\documentclass[12pt]{iopart}
\pdfoutput=1
\usepackage{color}
\usepackage{amssymb}
\usepackage{graphicx} 

\def\beqra{\begin{eqnarray}}
\def\eeqra{\end{eqnarray}}
\def\beq{\begin{equation}}
\def\eeq{\end{equation}}

\def\vp{\varphi}

\def\bx{{\bf{x}}}
\def\bk{{\bf{k}}}
\def\bp{{\bf{p}}}
\def\bq{{\bf{q}}}

\def\bv{{\bf{v}}}

\def\bV0{{\bf{V_0}}}
\def\re#1{(\ref{#1})}

\def\alt{\stackrel{<}{\sim}}
\def\bx{{\bf{x}}}
\def\by{{\bf{y}}}
\def\br{{\bf{r}}}

\def\bs{{\bf{s}}}
\def\bk{{\bf{k}}}
\def\bp{{\bf{p}}}
\def\bq{{\bf{q}}}

\def\bv{{\bf{v}}}

\def\bz{{\bf{z}}}

\begin{document}
\begin{flushright}
{\small UMN-TH-3439/15}
\end{flushright}

\title[The effect of massive neutrinos on the BAO peak]{The effect of massive neutrinos on the BAO peak}
\author{Marco Peloso$^{1}$, Massimo Pietroni$^{2,3}$, Matteo Viel$^{4,5}$, Francisco Villaescusa-Navarro$^{4,5}$}
\vskip 0.3 cm
\address{
$^1$School of Physics and Astronomy, University of Minnesota, Minneapolis, 55455, USA\\
$^2$INFN, Sezione di Padova, via Marzolo 8, I-35131, Padova, Italy\\
$^3$ Dipartimento di Fisica e Scienze della Terra, Universit\`a di Parma, Viale Usberti
7/A, I-43100 Parma, Italy\\
$^4$INAF - Osservatorio Astronomico di Trieste, Via G.B. Tiepolo 11, I-34143 Trieste, Italy\\
$^5$INFN, Sezione di Trieste, Via Valerio 2, I-34127 Trieste, Italy
}

\begin{abstract} 

We study the impact of neutrino masses on the shape and height of the BAO peak of the matter  correlation function, both in real and redshift space. In order to describe the nonlinear evolution of the BAO peak we run N-body simulations and compare them with simple analytic formulae. We show that the evolution with redshift of the correlation function and its dependence on the neutrino masses is well reproduced in a simplified version of the Zel'dovich approximation, in which the mode-coupling contribution to the power spectrum is neglected. 
While in linear theory the BAO peak decreases for increasing neutrino masses, the effect of nonlinear structure formation goes in the opposite direction, since the peak broadening by large scale flows is less effective. As a result of this combined effect, the peak decreases by  $\sim 0.6$ \% for  $ \sum m_\nu = 0.15$ eV and increases by $\sim1.2$\% for  $ \sum m_\nu = 0.3$ eV, with respect to a massless neutrino cosmology with equal value of the other  cosmological parameters.
 We extend our analysis to redshift space and to halos, and confirm the agreement between simulations and the analytic formulae.
 We argue that all analytical approaches having the Zel'dovich propagator in their lowest order approximation should give comparable performances, irrespectively to their formulation in Lagrangian or in Eulerian space.

\end{abstract}

\maketitle

The baryon acoustic oscillations (BAO) peak of the galaxy correlation function (CF) is one of the prominent observables in present day cosmology \cite{Eis05,Cole:2005sx}. The peak position corresponds to the acoustic scale at decoupling and can be used to infer the angular diameter distance as a function of redshift and the Hubble parameter $H(z)$. It is well known  that nonlinear effects of structure formation cause a  broadening of the peak, which degrades the measurement of the acoustic scale by roughly a factor of 3 at $z=0$ \cite{Eisenstein:2006nj}. The blurring of the peak is due mainly to the random displacement of the galaxies from their original location, by an average of $\sim 6\, {\mathrm{Mpc/h}}$. Reconstruction techniques have been developed \cite{Eisenstein:2006nk,Seo:2008yx,Padmanabhan:2008dd,Noh:2009bb,Tassev:2012hu,White:2015eaa} to undo the effect of these bulk motions and get back a CF closer to the linear one, and have been successfully applied to real data (see for instance \cite{Anderson:2013zyy}).
The effectiveness of these techniques, which are based essentially on the Zel'dovich approximation or on variants of it, can be seen as a proof, {\em a posteriori}, that the physics of the BAO peak degradation is well understood and under control.

 In this paper we explore the possibility of extracting cosmological information from these nonlinear effects, rather than seeing them as a noise to be removed. In practice, we will shift the focus from the peak position to the peak shape of the nonlinear CF, and we will study the latter as a function of redshift and of cosmological parameters, in particular neutrino masses.

Measuring the total mass of neutrinos by using cosmological data set is one of the most important goals of present and future large scale structure surveys. At present the tightest constraints are obtained by adding to the cosmic microwave background data, BAOs and the Lyman$-\alpha$ forest and result in a 2$\sigma$ C.L. upper limit of 0.14 eV \cite{Costanzi:2014tna,Palanque-Delabrouille:2014jca}. While BAOs are helpful in breaking degeneracies between the cosmological parameters probed by the cosmic microwave background (namely $\Omega_m$ and $H_0$), the Lyman-$\alpha$ forest data or other small scale structure probes like weak lensing and galaxy cluster data probe the neutrino induced suppression on the matter power spectrum. Although non zero neutrino masses have been claimed in the recent past in order to reconcile putative tensions between cosmic microwave background data and lower redshift probes, it appears that this solution is not supported by BAOs and the recent Planck results \cite{Planck_2015,Verde:2015ana}. In any case the tight upper limits above, if confirmed, will have tremendous implications for particle physics experiments \cite{DellOro:2015tia} and future cosmological searches like Euclid\footnote{http://www.euclid-ec.org}. These new findings are thus providing a strong motivitation to further explore neutrino induced non-linearities in unexplored regimes, like the one that we will be addressing in this paper.

\begin{figure}
\centering{ 
\includegraphics[width=.65\textwidth,clip]{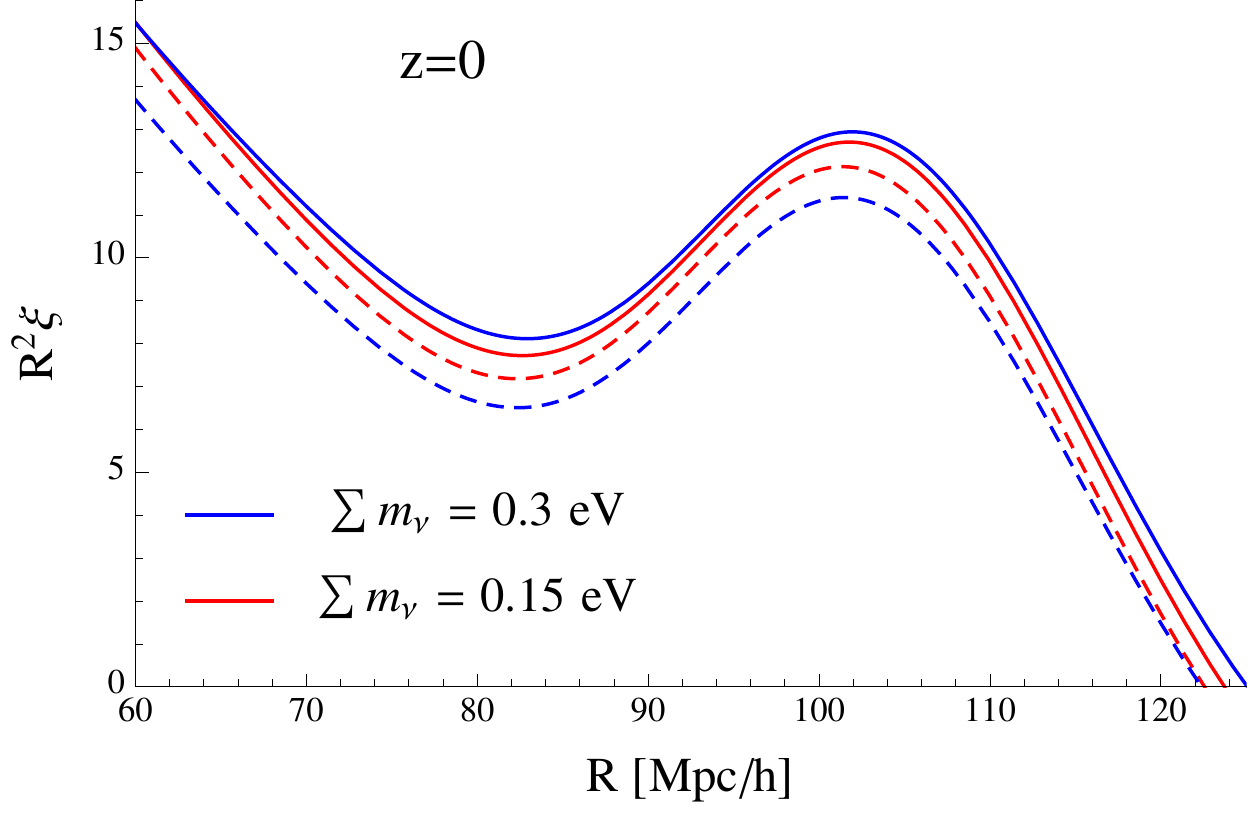}
}
\caption{Nonlinear  matter CFs in real space and at redshift $z=0$. The solid curves are for the massive neutrino cosmologies. The dashed curves are for cosmologies with massless neutrinos, but with the same values of $\Omega_b$, $\Omega_m$, and $\sigma_{8,{\rm cb}}$   as the corresponding cosmology with massive neutrinos. The CF has been computed as described in the text, and here and in all the other plots in this paper, it has been multiplied by $R^2$.
}
\label{figdegeneracy}
\end{figure}

Cosmologies with massive and massless neutrinos sharing the value of $\sigma_{8,{\rm cb}}$\footnote{This is the value of $\sigma_8$  computed using the CDM plus baryons power spectrum; in the following, $\Omega_m = 1 - \Omega_\Lambda$ denotes the present total dark matter density.} will have very similar halo mass functions \cite{Castorina:2013wga}. This implies that constraints on the sum of the neutrino masses from observables such as cluster number counts will be very degenerate with $\sigma_8$ (the well known $\Omega_\nu$-$\sigma_8$ degeneracy) \cite{Costanzi:2013bha}. One way to break that degeneracy is to use observables that prove larger scales, like the anisotropies in the cosmic microwave background or the galaxy correlation function on the BAO scale. In Figure \ref{figdegeneracy} we show the nonlinear CF for massive neutrino cosmologies (solid lines) and corresponding massless neutrino ones with the  same $\sigma_{8,{\rm cb}}$, $\Omega_b$, and $\Omega_m$ (dashed lines). As we see, the BAO peak provides a clearly distinct signature for these cosmologies. 

We will show that a very simple approximation to the nonlinear power spectrum (PS), consisting in removing the mode-coupling part from the Zel'dovich approximation, gives an accurate description of the BAO peak in the CF, and an even better one of  the ratio between CFs of different redshifts or of cosmologies with different neutrino masses. We also explore improved approximations, which give a better PS and CF, but do not improve substantially the CF ratios. Of course, the mode-coupling part is very important to recover the full nonlinear PS, where it becomes dominant at small scales, but since it is quite smooth on the BAO scales its contribution to the CF is limited to a subpercent shift in the acoustic scale, and almost disappears in CF ratios. 

We will consider the CF both for matter and halos, in real and in redshift space, and compare our analytic formulas to N-body simulations, finding perfect compatibility in all cases. 

The paper is organised as follows: in Section \ref{analytic} we discuss our analytic approximation to the CF in the BAO region, both in real and in redshift space, first for matter and then for halos; in Section \ref{NBODY} we present our suite of N-body simulations for $\Lambda$CDM and for massive neutrino cosmologies, and describe the algorithm to measure the CF; in Section \ref{comparison} we compare our analytic formulas with the CF measured from our simulations and also, for the $\Lambda$CDM cosmology, with the FrankenEmu \cite{Heitmann:2013bra} N-body based emulator; in Section \ref{conclusions} we summarize our results; in \ref{nlPS} we derive the general expression for the nonlinear PS, eq.~\re{PSnl}; in \ref{Zeldovich} we show the relation between our simplest approximation of the nonlinear CF, eq.~\re{PS1}, and the Zel'dovich approximation; in \ref{RSD} we derive the PS in redshift space including the effect of bulk flows. Finally, in \ref{sec:resolution_tests}, we discuss the dependence of our N-body simulations on mass and force resolution.

\section{Nonlinear evolution of the BAO peak}
\label{analytic}
In full generality, the nonlinear matter PS at redshift $z$  has the following structure \cite{RPTa,MP07b}:
\beq
P(k,z) = G^2(k,z) P^{lin}(k,z_{in}) + P_{MC}(k,z)\,,
\label{PSnl}
\eeq
where $P^{lin}(k,z)=D^2(z) P^{lin}(k,z_{in})$ is the linear PS, and $z_{in}$ is some initial redshift chosen well after decoupling and such that all the relevant scales are still in the linear regime (it can coincide with the redshift at which we start the N-body simulations, which in this paper is $z_{in}=99$, see Section \ref{NBODY}).  The nonlinear effects are completely encoded in the two functions appearing at the RHS: the {\em propagator} $G(k,z)$, representing the cross-correlator between the nonlinear density field at redshift $z$ and the initial one at $z_{in}$ \cite{RPTa,RPTb}, and the ``mode-coupling'' term $P_{MC}(k,z)$.

We stress that the above expression is completely general, the only assumption behind it being that the nonlinear density field, $\delta(\bk, z)$, is some ``functional'' of the initial density and velocity fields, see \ref{nlPS} for details. Then, one can compute these quantities in any consistent approximation scheme, such as Eulerian or Lagrangian perturbation theory (PT).

The BAO wiggles of $P^{lin}(k,z)$ are in general smoothed out in $P_{MC}(k,z)$ as  its computation involves momentum integrals in which two or more linear PS evaluated at different scales are convolved. Therefore, as we will demonstrate below, the BAO information is basically confined to the $G^2(k;z) P^{lin}(k)$ term which, after Fourier transform, accounts for approximately all the BAO peak in the CF. 

The propagator has been studied thoroughly in the recent literature \cite{RPTb,Anselmi:2010fs,Bernardeau:2011vy}. In Zel'dovich approximation it is given by (see \ref{Zeldovich})
\beq
G^{Zeld}(k,z) = \e ^{-\frac{k^2 \sigma_v^2(z)}{2}}\,,
\label{ZeldG}
\eeq
where $\sigma_v^2(z)$ is the 1-dimensional velocity dispersion evaluated in linear theory, namely,
\beq
\sigma_v^2(z) = \frac{1}{3}\int\frac{d^3 q}{(2\pi)^3}\,\frac{P^{lin}(q,z)}{q^2}\,.
\label{sigma2}
\eeq
In the following, we will mostly consider the CF obtained in Zel'dovich approximation by neglecting the mode-coupling part (see eq.~\re{Zeld3}). Namely, we will study the Fourier transform of 
\beq
P^{(1)}(k,z) =  e^{-{k^2 \sigma_v^2(z)}} P^{lin}(k,z) \,,
\label{PS1}
\eeq
and we will show that it gives a very good approximation to the nonlinear matter CF in the BAO peak region, and in particular to ratios of CF's for different redshifts, or for different values of neutrino masses, both in real and in redshift space. We will also show how it can be extended to describe halos.

In Figure \ref{fig:CF-matter-real-m00-z0z1}  we plot the linear CF, the CF obtained from \re{PS1}, and the CF from our N-body simulations for a $\Lambda$CDM cosmology (the corresponding list of cosmological parameters is given in Section \ref{NBODY}).  We also plot the difference between the  N-body based interpolator FrankenEmu \cite{Heitmann:2013bra} and the  CF obtained from \re{PS1}, namely  the CF associated to 
\beq
P(k,z) -P^{(1)}(k,z)=P_{MC}(k,z)\ +\left(  G^2(k,z)-  e^{-{k^2 \sigma_v^2(z)}} \right)P^{lin}(k,z) \,.
\label{Pexact-P1}
\eeq
As we anticipated, the simple formula \re{PS1} captures most of the physics responsible for the BAO peak broadening  ({\it i.e.} the difference between linear and nonlinear BAO peaks), while neglecting the fully nonlinear mode-coupling term and approximating the full propagator with its Zel'dovich limit gives negligible and mostly scale-independent contributions at these scales. 

As it has been emphasised in \cite{Senatore:2014via,Baldauf:2015xfa}, coherent bulk flows on scales larger than the BAO scale $l_{BAO}\simeq 110\, {\mathrm{Mpc/h}}$ cannot have any effect on the BAO peak in the CF. This is a manifestation of the general property of Extended Galilean Invariance (EGI) or of the Equivalence Principle, discussed in \cite{Scoccimarro:1995if,Peloso:2013zw,Peloso:2013spa,Kehagias:2013yd,Creminelli:2013mca}. Therefore, one would expect that the effect of modes $q\alt  2 \pi/l_{BAO}$ should be cut-off in the momentum integral in \re{sigma2}, in order to compute the CF at scales $O(l_{BAO})$ and smaller.  On the other hand,  the full expression \re{PSnl} automatically exhibits EGI:  the decoupling of the large modes is achieved via a cancellation of their effect between the propagator term, without any IR cutoff in the momentum integral, and the mode-coupling one.  The point is that, as we have seen, the mode-coupling term is smooth on the BAO peak scale, and therefore one can, as a first approximation, neglect it if one is interested in this feature. 

This approximation can be improved in two respects. On the one hand, the Zel'dovich approximation uses the linear expression for the displacement field at all scales, and will therefore fail for wavenumbers $q$ larger than some value $\Lambda$ in \re{sigma2}. On the other hand, eq.~\re{PS1} contains no mode-coupling effect at all, while these are known to induce some shift in the BAO peak position, although at a subpercent level \cite{Smith:2007gi,RPTa, Seo:2009fp, Padmanabhan:2009yr, Sherwin:2012nh}.

We can consider an improved version of our formula by using standard PT \re{PS1}, namely
\beqra
&&\!\!\!\!\!\!\!  \!\!\!\!\!\!\!\!  \!\!\!\!\!\!\!\! P^{(2)}(k,z) =  e^{-{k^2 \sigma_{v,\Lambda}^2(z)}} \left(P^{lin}(k,z)+P_{22}(k,z) + 2 \,P^{lin}(k,z) \Sigma_{\Lambda,\infty}(k,z)\right)\,,\label{impapprox}
\eeqra
where the 1-loop mode-coupling term is given by 
\beqra
&&\!\!\!\!\!\!\!\!\! \!\!\!\!\!\!\!\!\! \!\!\!\!\!\!\!\!\!  \!\!\!\!\!\!\!\!\!  P_{22}(k,z)=\nonumber\\
&& \!\!\!\!\!\!\!\!\! \!\!\!\!\!\!\!\!\!  \frac{k^2}{98\,(2\pi)^2} \int_0^\infty dq P^{lin}(q,z) \int_{-1}^1 dx P^{lin}(\sqrt{k^2+q^2-2 q k x},z) \frac{\left(3\frac{q}{k}+7x-10 \frac{q}{k}x^2\right)^2}{\left(1+\frac{q^2}{k^2}-2\frac{q}{k}x \right)^2}\,,\nonumber\\
&&
\eeqra
while 
\beqra
&&\!\!\!\!\!\!\!\!\!\!\!\!\!\!\! \!\!\! \!\!\! \!\!\! \!\!\! \!\!\! \!\!\! 
\Sigma_{\Lambda,\infty}(k,z)\equiv \frac{k^2}{504\, (2\pi)^2} \int_\Lambda^\infty dq\, P^{lin}(q,z)\Bigg[\frac{12 k^2}{q^2}\nonumber\\
&&\qquad -158+100 \,\frac{q^2}{k^2}-42\, \frac{q^4}{k^4}+\frac{3 k^3}{q^3} \bigg(\frac{q^2}{k^2}-1\bigg)^3\bigg(2+7 \frac{q^2}{k^2}\bigg) \log\left|\frac{k+q}{k-q}\right|\Bigg] \,,\nonumber\\
&&
\label{Sigma}
\eeqra
and $\sigma_{v,\Lambda}^2(z)$ is given by eq.~\re{sigma2} in which a UV cutoff on the momentum $q$ at the scale $\Lambda$ has been imposed. Using the fact that for soft modes $k^2 \sigma_{v,\Lambda}^2(z)\simeq -2 \,\Sigma_{0,\Lambda}(k,z)$, we see that 
expanding the exponential in \re{impapprox} one recovers the 1-loop expression for the PS in standard PT,
\beq
P^{(2)}(k,z) = P^{lin}(k,z)+P_{13}(k,z)+P_{22}(k,z) +O(\mathrm{2\,loop})\,.
\eeq

 The extension to higher loop orders, possibly including also a better treatment of the short modes along the effective approaches discussed in \cite{Manzotti:2014loa,Senatore:2014via,Baldauf:2015xfa},  is possible, although computationally more and more demanding. 
 
 Our main point in this paper is to show that the physics of the BAO peak degradation is well understood, so that by measuring not only the peak position, but also the peak shape, one can recover cosmological information. To discuss this point we will focus on the simplest approximation, eq.~\re{PS1}, keeping in mind that it can be systematically improved.

\subsection{Parametric dependence of the BAO peak}
We will concentrate on the BAO peak in the CF. We start by discussing pure matter in real space, for which the CF is given by
\beq
\xi(R) = \frac{1}{2 \pi^2 R} \int_0^\infty dq \,q  \sin(qR)\, P(q)\,,
\eeq
where we have omitted the time (or redshift) dependence. We will also use the ``moments'' $\xi_{1,2}(R)$, defined as
\beq
\xi_n(R) \equiv \frac{1}{2 \pi^2 R} \int_0^\infty dq \,q  \,(q R)^n\sin(qR)\, P(q)\,.
\eeq
In the BAO peak, $R=\bar{R}$, defined by 
\beq
\xi'(\bar R) = \frac{d \xi(\bar R)}{d \bar R}=0\,,
\eeq
we have $\frac{1}{2 \pi^2} \int_0^\infty d q \, q^2 \, \cos \left( q {\bar R} \right) \, P \left( q \right) = \xi \left( {\bar R} \right)$, and so 
 $\xi''(\bar R) = - \xi_2(\bar R)/\bar R^2$.

If we take the simplest expression \re{PS1} for $P(q)$, then the dependence of the CF $\xi(R)$ on $\sigma_v^2$ is given by
\beq
\frac{\partial \log \xi( R)}{\partial \log \sigma_v^2}= - \frac{\sigma_v^2}{R^2} \,\frac{\xi_2( R)}{\xi(R)}\,.
\eeq

If we consider two linear PS differing by $\delta P^{lin}(q)$, and, correspondingly, two $\sigma_v^2$ differing by
\beq
\delta \sigma_v^2 = \frac{1}{3}\int\frac{d^3 q}{(2\pi)^3} \frac{\delta P^{lin}(q)}{q^2}\,,
\eeq
the CF changes by 
\beqra
&&\delta \xi ( R) = \frac{1}{2 \pi^2 R} \int dq\,q \,\sin(q R) \;\delta P^{lin}(q) \,e^{-q^2 \sigma_v^2}
- \frac{\xi_2 \left( R \right) \, d \sigma_v^2}{R^2} \nonumber\\
&&\qquad \;=  \frac{1}{2 \pi^2} \int dq\,q^2\;\delta P^{lin}(q)\left( \frac{\sin (q R)}{q  R} \,e^{-q^2 \sigma_v^2}-\frac{1}{3}\frac{ \xi_2( R)}{q^2 R^2}\right)\,.
\label{shift}
\eeqra

For instance, if the two linear PS differ only in their normalizations,   $\delta P^{lin}(q) = P^{lin}(q) \delta A/A $, the corresponding height of the BAO peak changes by
\beq
\frac {\delta_A \xi ( R)}{ \xi ( R)} = \frac{\delta A}{A}\left( 1- \frac{\sigma_v^2 \, \xi_2( R)}{R^2  \xi( R)} \right)\,,
\eeq 
where the second term inside parentheses gives the --scale-dependent-- nonlinear effect.
A special case is that of considering two different (and nearby) redshifts, in which case $\delta A/A =\delta D(z)^2/D(z)^2$. In this case, 
at $z=0$, the nonlinear term gives typically a $~15-25\%$ negative correction at the peak position and a $~20\%$ positive one at the minimum left to the BAO peak, with respect to the linear result.

Eq.~\re{shift} can also be used to estimate the peak change between two slightly different cosmologies. For instance, let us consider the change of the BAO peak at $z=0$ between two cosmologies with different neutrino masses (the two cosmologies only differ from each other by the neutrino mass, and by the cold dark matter abundance, in such a way that  the total matter component, $\Omega_{\rm m}$, is the same). For $ \sum m_\nu = 0.15$ eV, eq.~(\ref{shift}) predicts a decrease of the peak height of $\sim - 0.6 \%$ with respect to the massless neutrino case. At the higher masses that we have considered  the peak heights is instead greater than in the massless case.  Specifically, eq. \re{shift} gives an increase of $\sim 1.2\%$  for $\sum m_\nu = 0.3$ eV, and of 
$\sim 5.7 \%$ in the case of  $\sum m_\nu = 0.6$ eV. This behavior is confirmed by the results presented in Figure  \ref{fig:CF-matter-real+redshift-ratm}. In particular, we see  that nonlinear effects invert the trend with respect to linear theory. Indeed, in linear theory the BAO peak height  decreases with  increasing neutrino masses (see the red-dashed curves at $R \sim 100 \, {\rm Mpc}/ h$ in the figure). The bulk flows, on the other hand, are less effective for higher neutrino masses (as the corresponding $\sigma_v^2$ is lower) in degrading the linear BAO peak, and therefore they increase the ratio of the height in the massive vs. massless case with respect to the linear prediction.  For $\sum m_\nu = 0.15$ eV, the decrease of the peak predicted by the linear theory dominates over the increase due to the bulk flows.  The opposite is true for  $\sum m_\nu = 0.3 ,\, 0.6$ eV.

In Section \ref{comparison} we compute the CF in real and redshift space, starting from the power spectra introduced above. Specifically, in real space we compute and show 
\begin{eqnarray}
\xi^{\rm lin} \left( R  \right) &=& \frac{1}{2 \pi^2 R} \int_0^\infty d q \, q \, \sin \left( q R \right) P^{\rm lin} \left( q  \right) \,, \nonumber\\ 
\xi^{(1)} \left( R  \right) &=& \frac{1}{2 \pi^2 R} \int_0^\infty d q \, q \, \sin \left( q R \right) {\rm e}^{- q^2 \, \sigma_v^2 \left( z \right)} P^{\rm lin} \left( q \right) \,, \nonumber\\ 
\xi^{(2)} \left( R  \right) &=& \frac{1}{2 \pi^2 R} \int_0^\infty d q \, q \, \sin \left( q R \right) {\rm e}^{- q^2 \, \sigma_v^2 \left( z \right)} \left[ P^{\rm lin} \left( q \right) + P_{22} \left( q  \right) \right] \,,  
\label{zeta-lin-1-2}
\end{eqnarray}
where the last one is obtained by taking the $\Lambda \rightarrow \infty$ limit in eq. (\ref{impapprox}). As we will see, $\xi^{(2)}$ reproduces the shape and position of the BAO peak slightly better than  $\xi^{(1)}$. However, they perform equally well in reproducing ratios of the CF between different redshifts (within a fix cosmology) or  ratios of correlations functions of different cosmologies (at equal redshift). For this reason, we only use  $\xi^{(1)}$ in the subsequent analysis. 

Specifically, after studying the matter CF in real space, we study the matter CF in redshift space, and the CF for biased halos that are predicted from $P^{(1)}$, and we compare them with the results of the numerical simulations. We verify that the agreement is comparable with that obtained for the matter CF in real space. 

\subsection{Redshift space}
In redshift space, the PS is a function of time, of the magnitude $k$ of the momentum of the modes, and of the cosine $\mu=\hat \bk\cdot \hat \bz$ of the angle between the line of sight (that we take along the $z$-axis)  and $\bk$. On large scales the PS is given by the Kaiser formula \cite{Kaiser:1987qv}
\begin{equation}
P_s^{\rm Kaiser} \left( k ,\, \mu  \right) = \left( 1 + \mu^2 \, f \right)^2 P^{\rm lin} \left( k \right) \,,
\label{kaiser}
\end{equation}
where $f = \frac{d \, \log D}{d \log a}$ and $a$ is the scale factor (to simplify the notation, the time dependence of $f$ is understood; in the following, from now on, also the time dependence of $\sigma_v^2$ is understood). The same effect on the nonlinear PS  (\ref{PS1})  gives 
\begin{equation}
P_s^{(1)} \left( k ,\, \mu  \right) = P_s^{\rm Kaiser} \left( k ,\, \mu  \right)  \, {\rm e}^{-k^2 \sigma_v^2 \left[ 1 + \mu^2 f \left( 2 + f \right) \right]} \,,
\label{PS1-redshift}
\end{equation}
see \ref{RSD}.

The CF has also nontrivial angular dependence, $\xi_s \left( {\bf R}  \right)$. In the following, we only study its  monopole component. For a generic PS that is function of $k$ and $\mu^2$, the angular averaged CF is 
\begin{equation}
\!\!\!\!\!\!\!\! \!\!\!\!\!\!\!\! 
{\bar \xi}_s \left( R  \right) \equiv \frac{1}{4 \pi} \int d \Omega_{\hat R} \, \xi_s \left( {\bf R}  \right) 
= \frac{1}{2 \pi^2 R} \int_0^\infty d q \, q \, \sin \left( q R \right) \int_0^1 d \mu \, P_s \left( k ,\, \mu \right) \,. 
\label{general-zeta-mu}
\end{equation} 
Using the approximations \re{kaiser} and \re{PS1-redshift} in \re{general-zeta-mu} we get, respectively, 
\begin{eqnarray}
&& \!\!\!\!\!\!\!\! \!\!\!\!\!\!\!\! \!\!\!\!\!\!\!\! 
{\bar \xi}_s^{\rm Kaiser} \left( R \right) = \left( 1 + \frac{2}{3} f + \frac{f^2}{5} \right) \, \xi^{\rm lin} \left( R  \right) \,, 
\nonumber\\ 
&& \!\!\!\!\!\!\!\! \!\!\!\!\!\!\!\! \!\!\!\!\!\!\!\! 
{\bar \xi}_s^{(1)} \left( R  \right) = \frac{1}{2 \pi^2 R} \, \int_0^\infty d q \, q \, \sin \left( q R \right) P^{\rm lin} \left( q \right)  {\rm e}^{- q^2 \sigma_v^2 } {\cal F} \left[ \sqrt{2 + f} \, q \, \sigma_v ,\: f ,\: 1 \right] \,, \nonumber\\ 
\label{kaiser-1}
\end{eqnarray} 
where we have defined 
\begin{equation}
\!\!\!\!\!\!\!\! \!\!\!\!\!\!\!\! \!\!\!\!\!\!\!\! \!\!\!\!\!\!\!\! 
{\cal F} \left[ K ,\: f ,\: A \right] \equiv 
\frac{ 3 + 4 \, K^2 \left( A +  A^2 \, K^2 \right) }{8 K^4} \, \frac{\sqrt{\pi} \, {\rm Erf} \left[ \sqrt{f } \, K \right]}{\sqrt{f } K} 
 - \frac{  3 + 2 \left( 2 \, A + f \right) K^2  }{4 K^4} \, {\rm e}^{- f  K^2} \,. 
\label{calF}
\end{equation} 

\subsection{Halos}
We are also interested in CFs of dark matter halos. Following \cite{Desjacques:2008jj,Desjacques:2009kt,Baldauf:2014fza}, we consider the following large scale bias predicted by the peaks model between the halo and the matter density fields 
\begin{equation}
\frac{\left\langle \delta_m^{\rm lin} \left( k, z \right)  \delta_h \left( k, z \right) \right\rangle}{
\left\langle \delta_m^{\rm lin} \left( k, z \right)  \delta_m^{\rm lin} \left( k, z \right) \right\rangle}
= \left[ b_{10} \left( z \right) + b_{01} \left( z \right) \, k^2 \right] \, {\rm e}^{- k^2 \sigma_v^2 \left( z \right) / 2 } \,, 
\label{fit-bias}
\end{equation}
where, in addition to what done in  \cite{Desjacques:2008jj,Desjacques:2009kt,Baldauf:2014fza}, we have added the exponential factor to account for the (unbiased) bulk flow suppression.  We extract the coefficients $b_{10} ,\, b_{01}$ by matching  at largest scales the correlation function obtained from our  N-body simulations. 

From this relation, we obtain the real space halo-halo CF
\begin{equation}
\xi_{hh}^{(1)} \left( R  \right) = \frac{1}{2 \pi^2 R} \int_0^\infty d q \, q \, \sin \left( q R \right) \left[ b_{10} + b_{01} \, q^2 \right]^2 
 {\rm e}^{- q^2 \, \sigma_v^2} P^{\rm lin} \left( q \right) \,. 
\label{CF1-hh}
\end{equation}

Strictly speaking, this relation is only valid on large scales, but we show that assuming it to be valid at all scales allows for a good reconstruction of the BAO peak in the halo CF: the measurement of the halo CF from the N body simulations has a greater uncertainty than that of matter, due to the smaller statistics (there are fewer halos than dark matter particles in the simulation; this is particularly true at higher redshift, where the halo data become progressively more uncertain). 
We find that the CF (\ref{CF1-hh}) agrees with the numerical simulations as well as in the matter case, compatible with the increased error bars. 

 We notice that using N-body simulations it was found by \cite{Villaescusa-Navarro:2013pva, Castorina:2013wga} that the halo bias in cosmologies with massive neutrinos is scale dependent, even on very large scales. The reason for this is that dark matter halos are biased tracers on the underlying CDM+baryons field, not of the total matter one. If the halo bias is defined with respect to the underlying CDM+baryon field, the halo bias turns out to be scale independent on large scales and universal, as found by \cite{Castorina:2013wga}. Therefore, in cosmologies with massive neutrinos it is better to compute the halos correlation function using the above equation but employing the properties of the CDM+baryons field rather than these of the total matter. Given the fact that the error bars from our halo catalogue are very large, and since the scale-dependence bias effect induced by massive neutrinos is small, we decided, for simplicity and clearness, to employ the above equation even for cosmologies with massive neutrinos.

Finally, we are interested in the halo CF in redshift space. We start from (\ref{CF1-hh}), and assume that the halo velocity is unbiased, so to obtain the redshift space PS 
\begin{equation}
P_{s,hh}^{(1)} \left(k  \right)= \left[ \left( b_{10} + b_{01} k^2 \right) + \mu^2 f \right]^2 P^{\rm lin} \left( k \right) \, {\rm e}^{-k^2 \sigma_v^2  \left[ 1 + \mu^2 f \left( 2 + f \right) \right]} \,. 
\label{PS-shh}
\end{equation} 
Using (\ref{general-zeta-mu}), we obtain 
\begin{eqnarray} 
\!\!\!\!\!\!\!\! \!\!\!\!\!\!\!\! \!\!\!\!\!\!\!\! \!\!\!\!\!\!\!\! 
{\bar \xi}_{s,hh}^{(1)} \left( R  \right) &=& \frac{1}{2 \pi^2 R} \, \int_0^\infty d q \, q \, \sin \left( q R \right) P^{\rm lin} \left( q  \right)  {\rm e}^{- q^2 \sigma_v^2 } {\cal F} \left[ \sqrt{2 + f} \, q \, \sigma_v ,\, f , b_{10} + q^2 \, b_{01} \right] \,, \nonumber\\ 
\label{CF1-hh-redshift}
\end{eqnarray}
where ${\cal F}$ has been defined in (\ref{calF}).

\section{N-Body simulations}
\label{NBODY}
We have run N-body simulations for four different cosmological models; one with massless neutrinos and three with massive neutrinos: $\sum m_\nu=0.15$ eV, $\sum m_\nu=0.3$ eV and $\sum m_\nu=0.6$ eV. All simulations share the value of the following cosmological parameters, which are in excellent agreement with the latest results by the Planck collaboration \cite{Planck_2015}: $\Omega_{\rm m}=0.3175$, $\Omega_{\rm b}=0.049$, $\Omega_\Lambda=0.6825$, $h=0.6711$, $n_s=0.9624$, $A_s=2.13\times10^{-9}$. In cosmologies with massive neutrino we set the value of $\Omega_{\rm cdm}$ to $\Omega_{\rm m}-\Omega_{\rm b}-\Omega_\nu$, where $\Omega_\nu h^2=\sum m_\nu/(94.1~{\rm eV})$. The value of $\sigma_8$ obtained from these parameters is 0.834, 0.801, 0.764, 0.693 for the models with $\sum m_\nu=0.0$, 0.15, 0.3 and 0.6 eV neutrinos, respectively. 
 
For each cosmology we have run 100 independent N-body simulations~\footnote[1]{ We associate an error to the mean two-point correlation function equal to the variance around the mean from the 100 realizations run for every cosmological model divided by $\sqrt{10}$.} using the TreePM code {\sc GADGET-III} \cite{Gadget_II}. In each realization we follow the evolution of $256^3$ cold dark matter (plus $256^3$ neutrino particles for massive neutrino models) down to $z=0$ in a periodic box of 1000 Mpc$/h$ length. The Plummer equivalent gravitational softening is set, for each particle type, to 1/40 of the mean inter-particle linear spacing. We save snapshots at redshifts 2, 1, 0.5 and 0.

We have generated the N-body initial conditions, at $z_{in}=99$, in the following way: we set the particles in a regular cubic grid and then we displace their positions according to the Zel'dovich approximation. The linear matter power spectra and transfer functions are computed using the CAMB code \cite{CAMB}. In simulations with massive neutrinos we use a transfer function which is a mass-weighted average between the CDM and baryons transfer functions to set up the initial positions and velocities of the CDM particles.

Dark matter halos are identified using the Friends-of-Friends algorithm \cite{FoF} taking a value for the linking length parameter equal to $b=0.2$. We only consider halos containing at least 32 CDM particles, i.e. halos with masses larger than $\sim1.6\times10^{14}~h^{-1}{\rm M}_\odot$  The two-point correlation function of the dark matter halos is computed using the Landy-Szalay estimator \cite{Landy_Szalay}
\begin{equation}
\xi(r)=\frac{DD(r)-2DR(r)+RR(r)}{RR(r)}
\end{equation}
with $DD(r)$ and $RR(r)$ being the number of pairs with distances in the interval $[r-\triangle r,r+\triangle r]$ in the halo and random catalogues, respectively. $DR(r)$ is the number of halo-random points pairs in the above interval. Our random catalogue contains half million points, which is a number more than 30 times larger than the typical number of halos in one realization at $z=0$. 

Unfortunately, the use of the Landy-Szalay estimator to compute the two-point correlation function of the matter field is computationally unfeasible, given the large number of tracers in our simulations. Therefore, we evaluate the matter correlation function using the estimator presented in \cite{Taruya:2009ir} that we now briefly describe. We first evaluate the density contrast field, $\delta(\vec{r})$, using a grid with $256^3$ (or $512^3$) cells using the cloud-in-cell (CIC) mass assignment scheme. We then Fourier transform $\delta(\br)$ to obtain $\delta(\bk)$ and compute in each point of the grid the quantity $|\delta(\bk)|^2$. Finally, we Fourier transform the previous quantity and compute the two-point correlation function by averaging over all modes that lie within the same $r-$bin. We set the width of the bins in radius equal to $L/N_{\rm cells}^{1/3}$, where $L$ is the size of the simulation box. 

In order to verify the robustness of the above algorithm we have also computed the two-point correlation function using the estimator presented in \cite{Barriga_Gaztanaga,Eriksen_2003, Sanchez_2008}.~\footnote{ We first compute the density field in a grid with $160^3$ cells using the nearest grid point mass assignment scheme and then we evaluate the correlation function using the estimator 
$\xi(r)=\frac{1}{N_{\rm pairs}}\sum_{i,j}\delta_i \delta_j~,$
where the sum is performed over cells whose distance lie in the interval $[r-\triangle r,r+\triangle r]$ and $N_{\rm pairs}$ is the number of cell pairs in that interval.} We find an excellent agreement among the correlation functions computed using the two different estimators. We notice that matter correlation function results shown in the paper have been obtained using the former estimator.

Correlation functions in redshift-space are computed using the above methods after displacing the particle (or dark matter halo centers) positions from real-space to redshift-space using eq.~\re{rsshift}.
We have created three different catalogues, by displacing the particles/halos along each cartesian axis according to the peculiar velocities projected along that axis, and in the paper we show the results when averaging among the three.

\section{Results}
\label{comparison}

In this section we discuss the dependence of the CF on redshift and on values of the neutrino masses, and compare the results of N-body simulations to the analytic formulas discussed in Sect.~\ref{analytic}. 

\begin{figure}
\centering{ 
\includegraphics[width=.45\textwidth,clip]{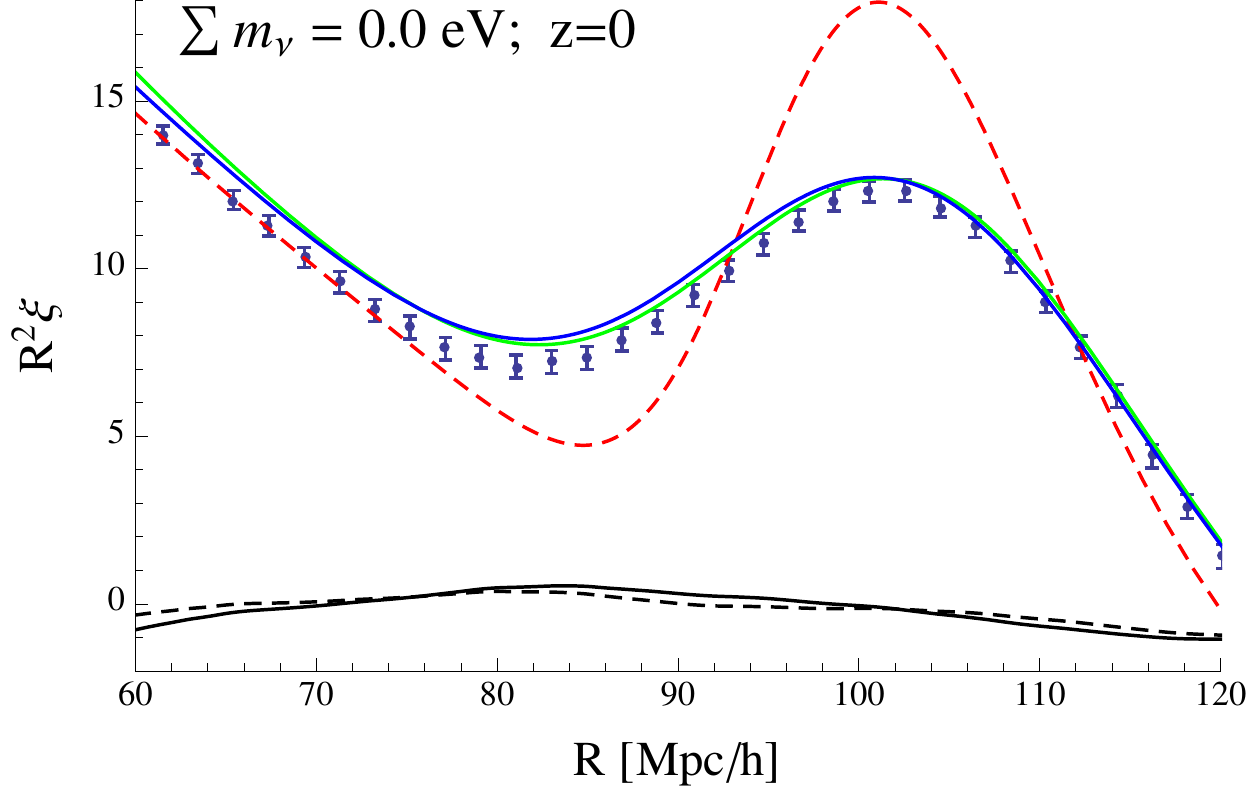}
\includegraphics[width=.45\textwidth,clip]{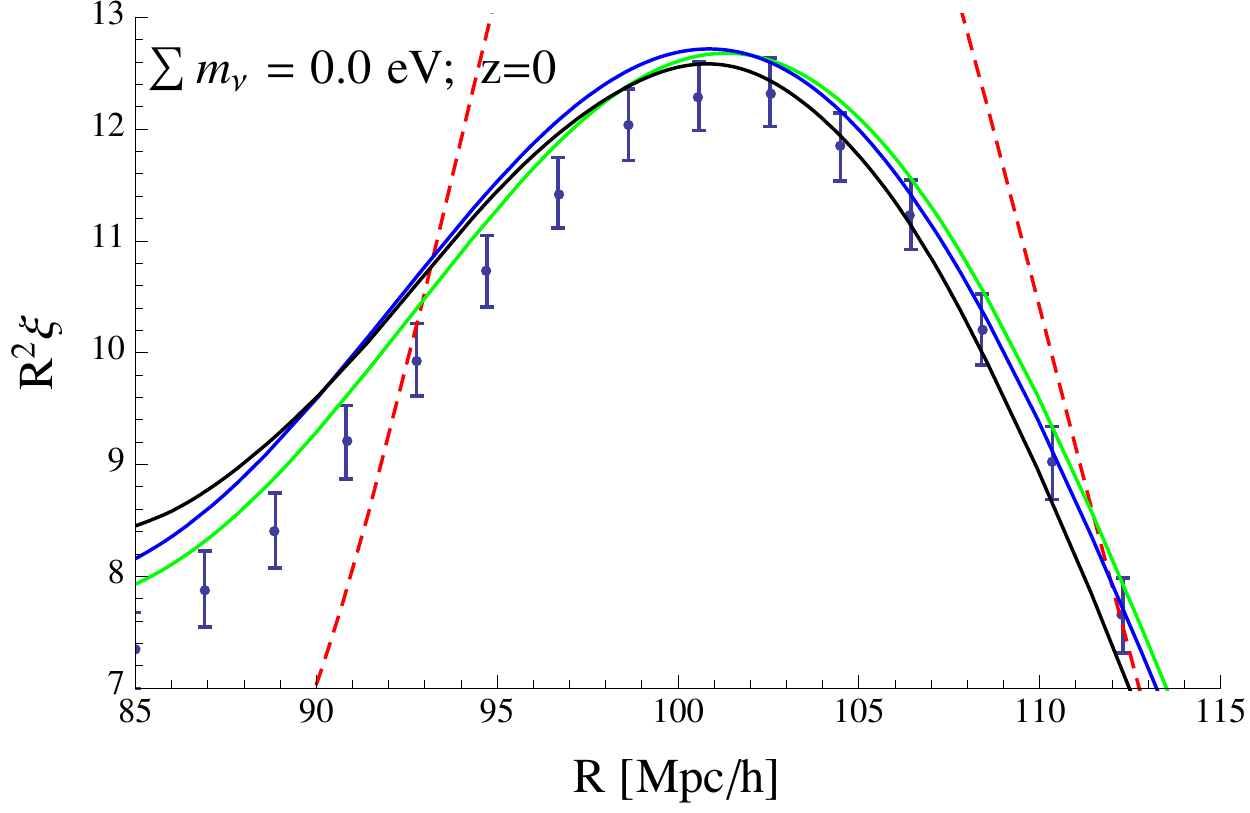}
}
\centering{ 
\includegraphics[width=.45\textwidth,clip]{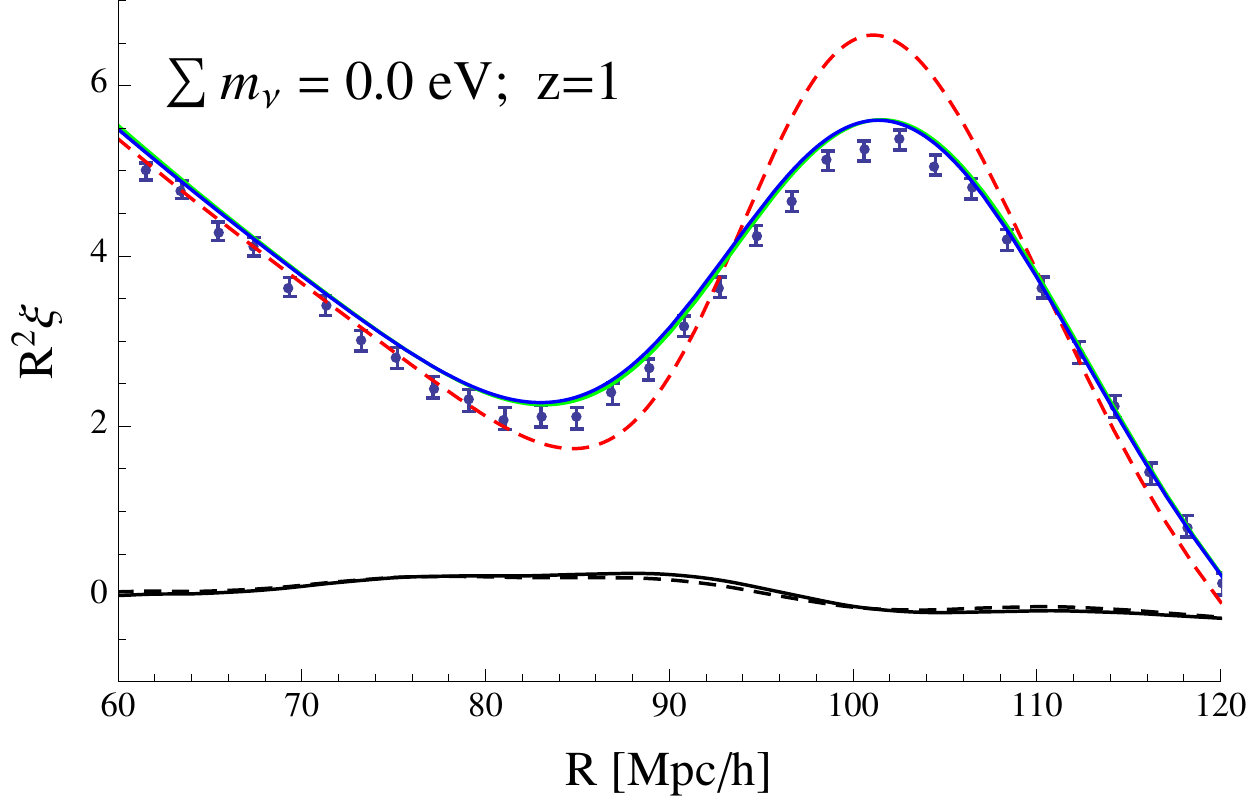}
\includegraphics[width=.45\textwidth,clip]{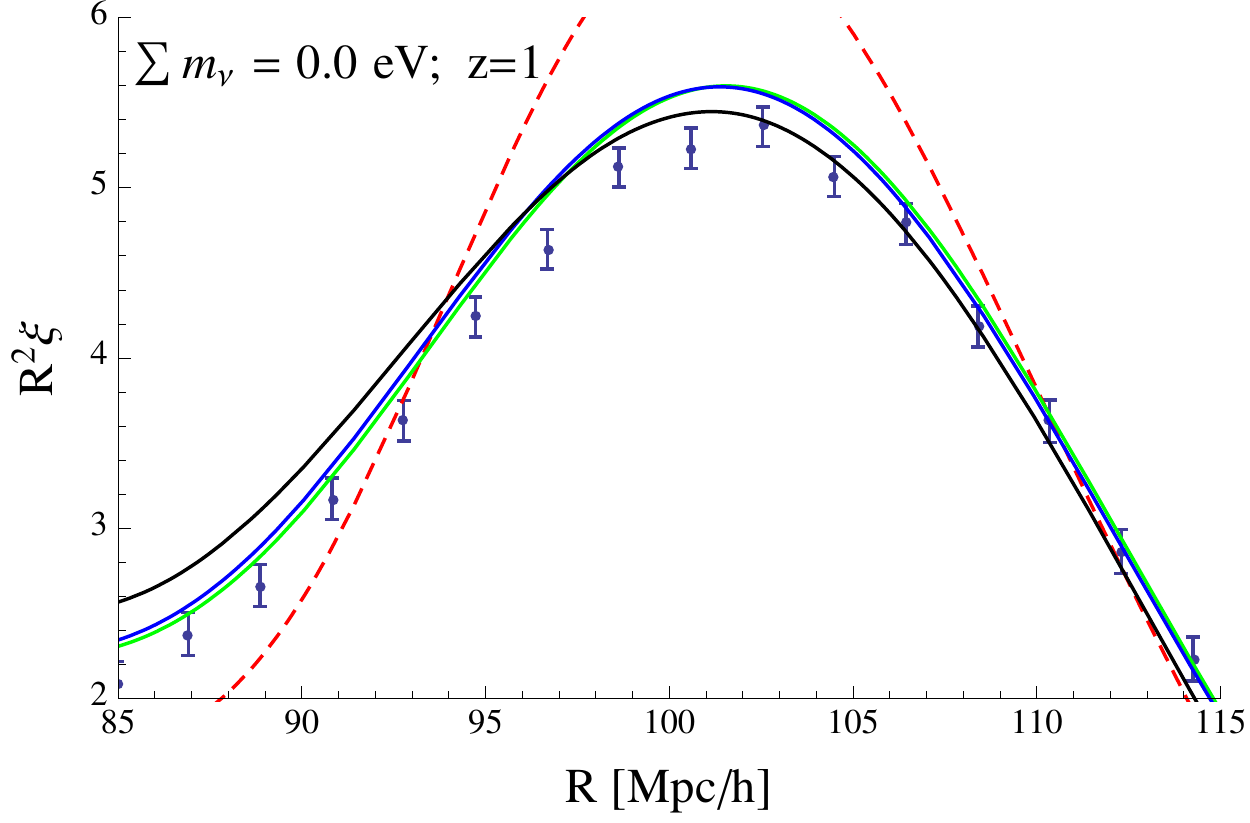}
}
\caption{\label{fig:CF-matter-real-m00-z0z1} First row: Matter CF in real space, for massless neutrinos, and at redshift $z=0$. The right panel is a zoom of the left panel centered at the BAO peak.  The data points are from our N-body simulations; the red dashed, green solid, and blue solid lines are, respectively, $\xi^{\rm lin} ,\, \xi^{(1)}, {\rm and } \, \xi^{(2)}$, defined in eq. (\ref{zeta-lin-1-2}), multiplied by $R^2$. The black solid (dashed) line at small $R^2 \xi$ values in the left panel is the difference (\ref{Pexact-P1}) between the CF from the  FrankenEmu \cite{Heitmann:2013bra} N-body based emulator and $\xi^{(1)}$ (and  $\xi^{(2)}$), also rescaled by $R^2$. The black solid line in the right panel is the  FrankenEmu  CF, times $R^2$. Second row: same as in the first row, but at redshift $z=1$. 
}
\end{figure}

In Figure \ref{fig:CF-matter-real-m00-z0z1} we show the matter CF in the massless neutrino case at $z=0$ (first row) and at $z=1$ (second row). The right panels are a zoom of the left panels centered at the BAO peak. We see that linear theory predicts a substantially higher BAO peak, and we clearly see the nonlinear broadening in $\xi^{(1)}$ and  $\xi^{(2)}$.   In the left panels of the figure, we also show with a black solid (dashed) line the residual difference between the CF obtained from the FrankenEmu \cite{Heitmann:2013bra} N-body based emulator, and $\xi^{(1)}$ (and $\xi^{(2)}$). This difference -- which for  $\xi^{(1)}$ is mostly due to the mode-coupling contribution to the PS -- appears to be very weakly dependent on $R$ and on the redshift. We stress that the exponential factor in \re{PS1} is not fitted to the data, but it is obtained from the linear PS, according to \re{sigma2}.

\begin{figure}
\centering{ 
\includegraphics[width=.65\textwidth,clip]{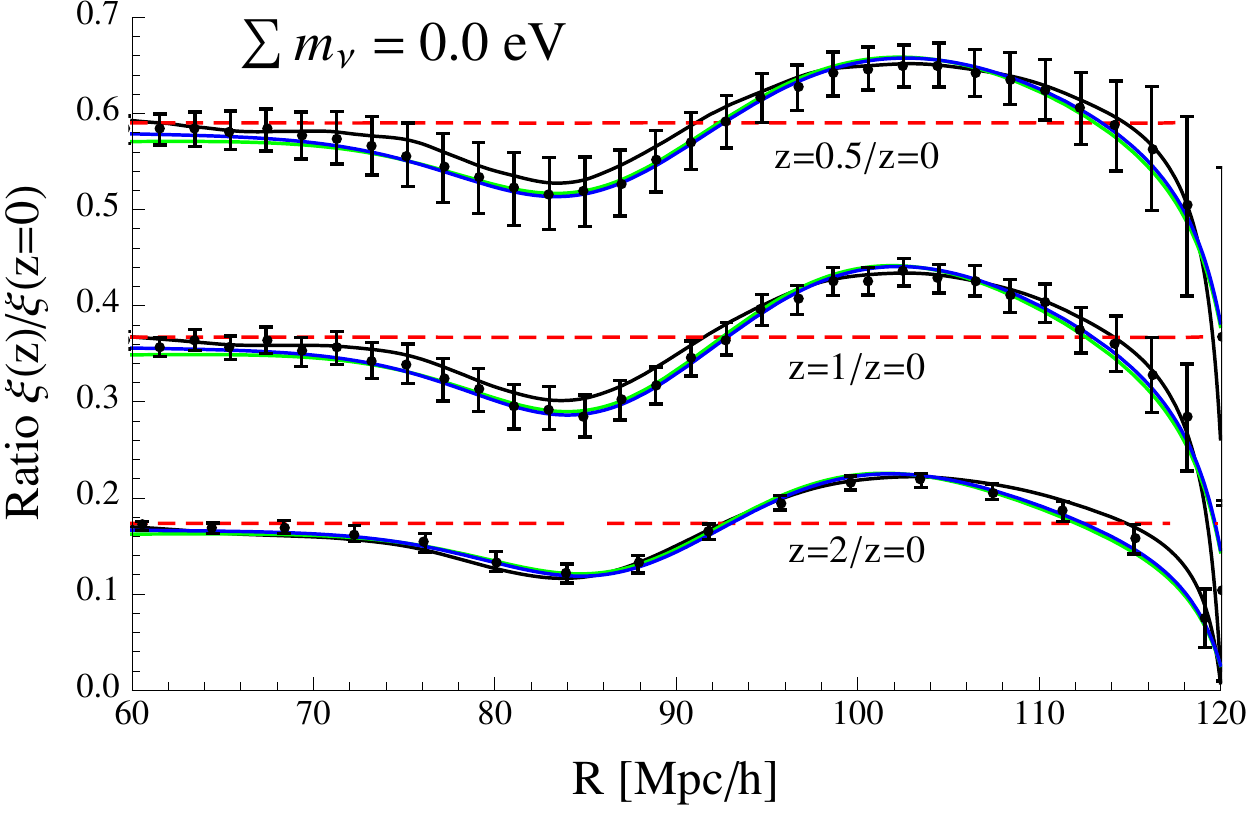}
}
\caption{\label{fig:CF-matter-real-m00-zrat} Ratio between the matter real space CF at two different redshifts, for massless neutrinos. The top, middle, and bottom curves in the figure are ratios of CF at $z=0.5,\, z=1,\, z=2$, respectively, divided by the corresponding CF at $z=0$. The data are ratios between our N-body simulations;  the red dashed, green solid, and blue solid lines are ratios between, respectively, $\xi^{\rm lin} ,\, \xi^{(1)}, {\rm and } \, \xi^{(2)}$,  defined in eq. (\ref{zeta-lin-1-2}). The black solid lines are ratios between CF obtained from  the FrankenEmu emulator. }
\end{figure}

\begin{figure}
\centering{ 
\includegraphics[width=.45\textwidth,clip]{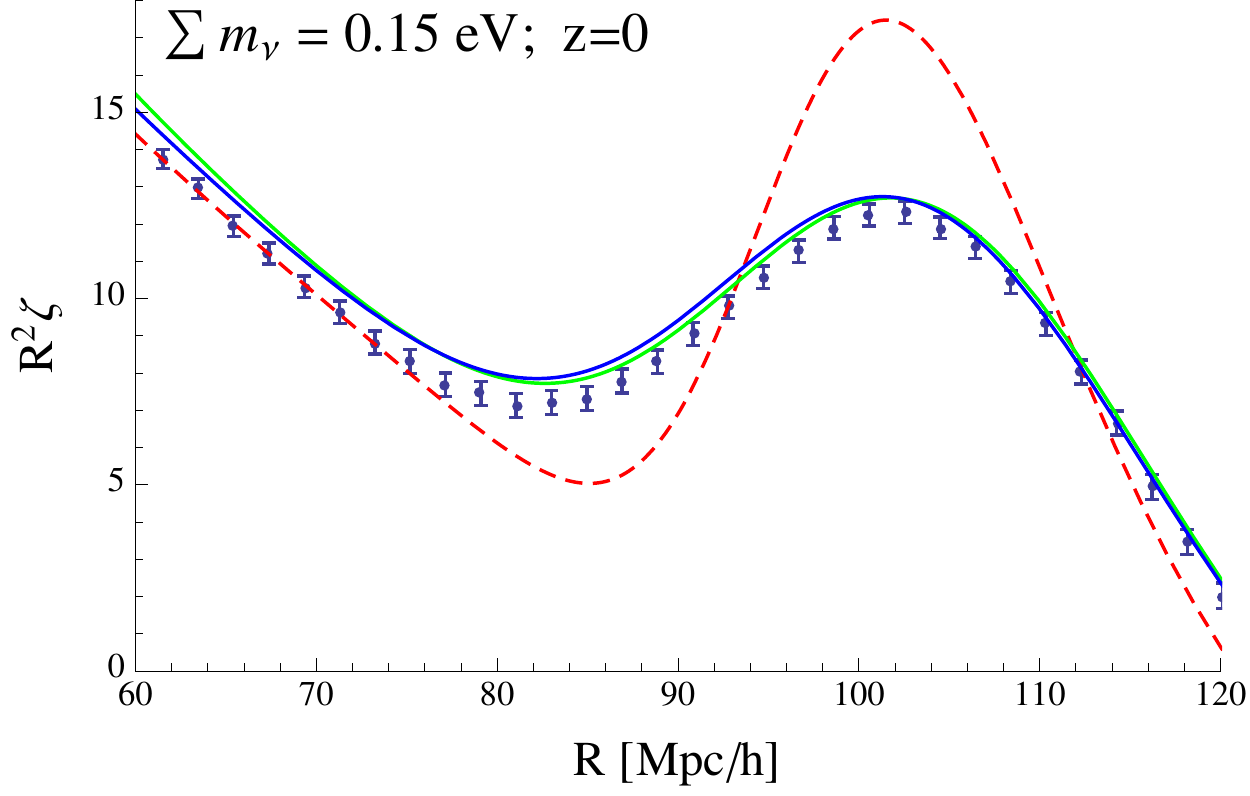}
\includegraphics[width=.45\textwidth,clip]{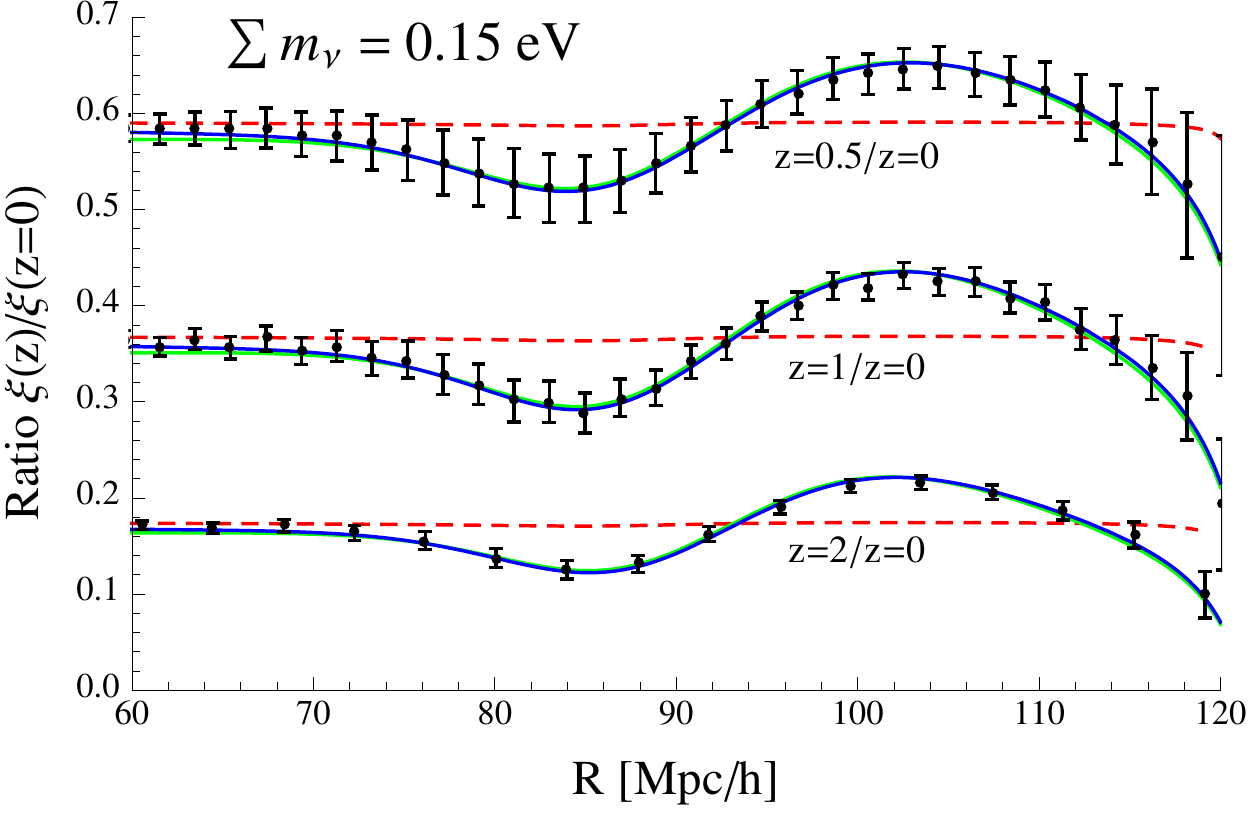}
}
\centering{ 
\includegraphics[width=.45\textwidth,clip]{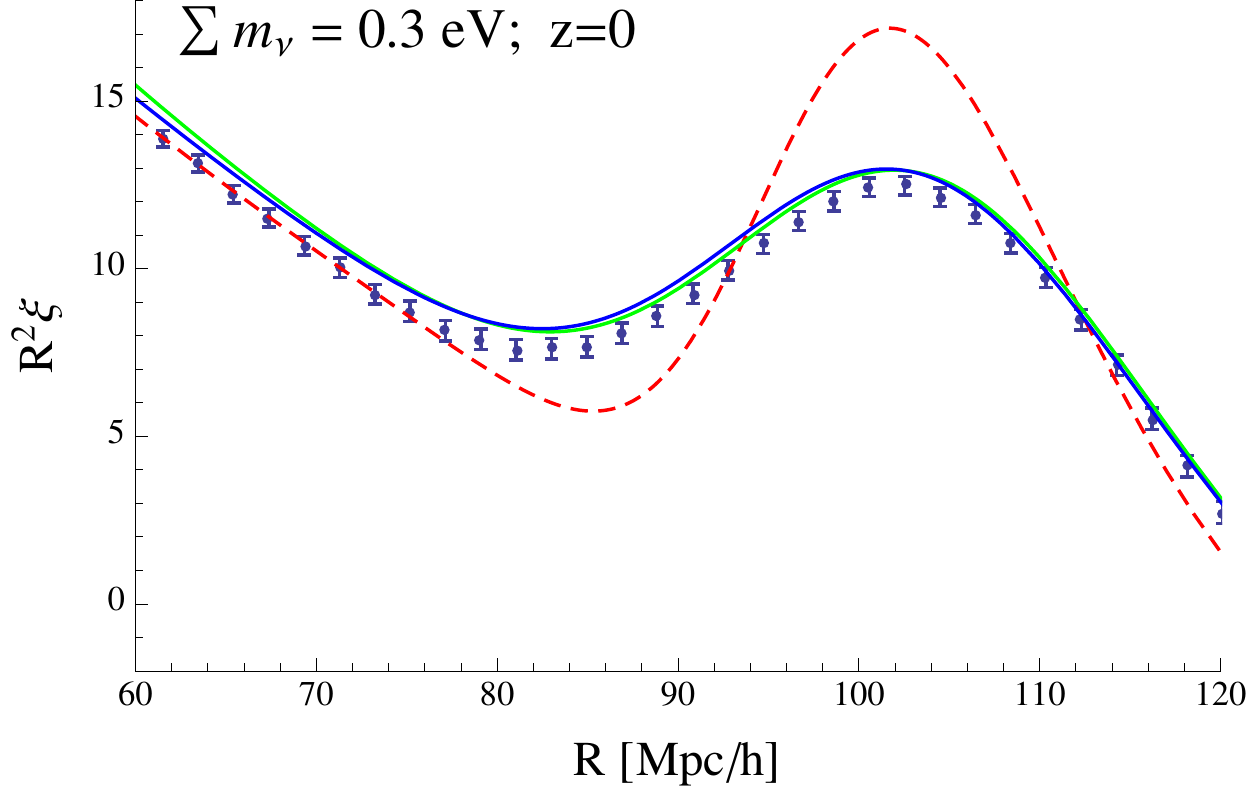}
\includegraphics[width=.45\textwidth,clip]{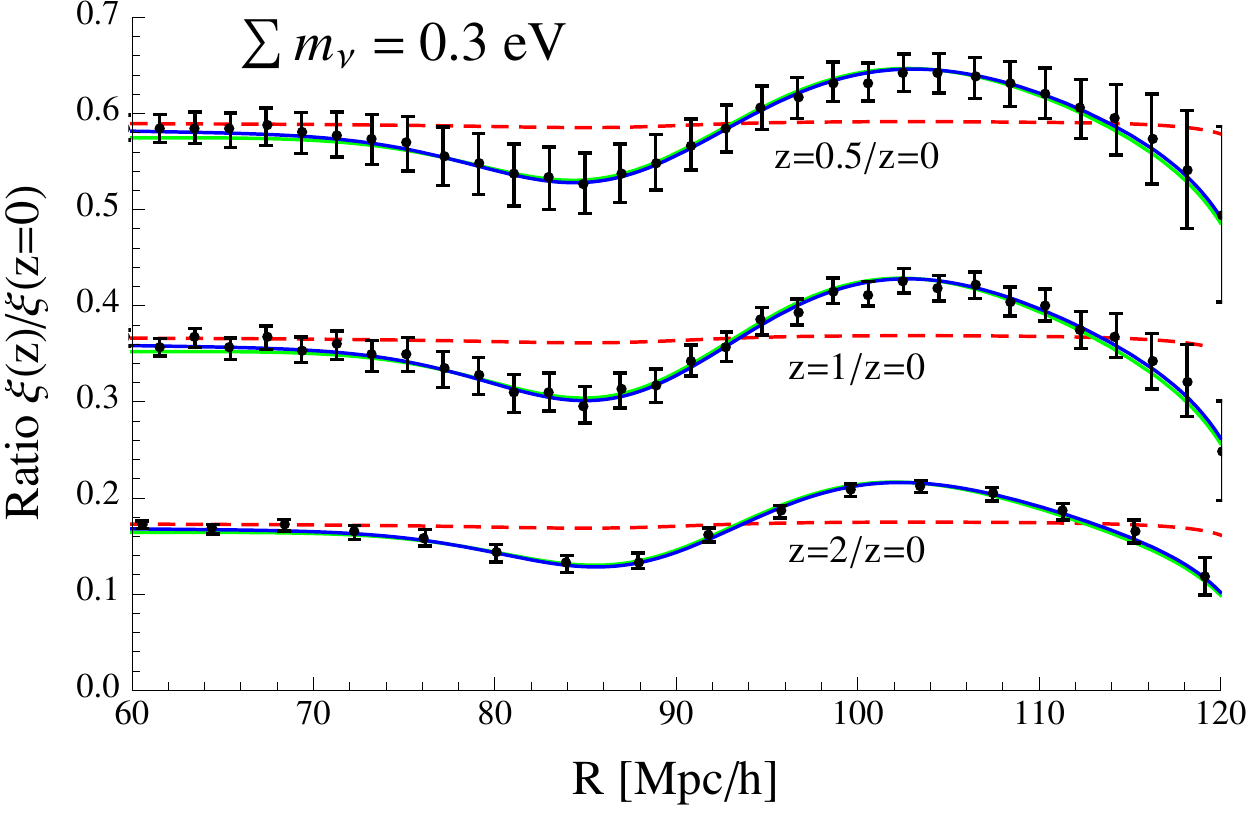}
}
\caption{\label{fig:CF-matter-real-m015-m03} Matter CFs in real space. Analogous of Figures \ref{fig:CF-matter-real-m00-z0z1}  ($z=0$) and \ref{fig:CF-matter-real-m00-zrat} (ratios between CFs at different $z$), but now for massive neutrinos. The figures in the first row are for $\sum m_\nu = 0.15 \, {\rm eV}$, while those in the second row are for $m_\nu = 0.3$ eV.  }
\end{figure}

\begin{figure}
\centering{ 
\includegraphics[width=.45\textwidth,clip]{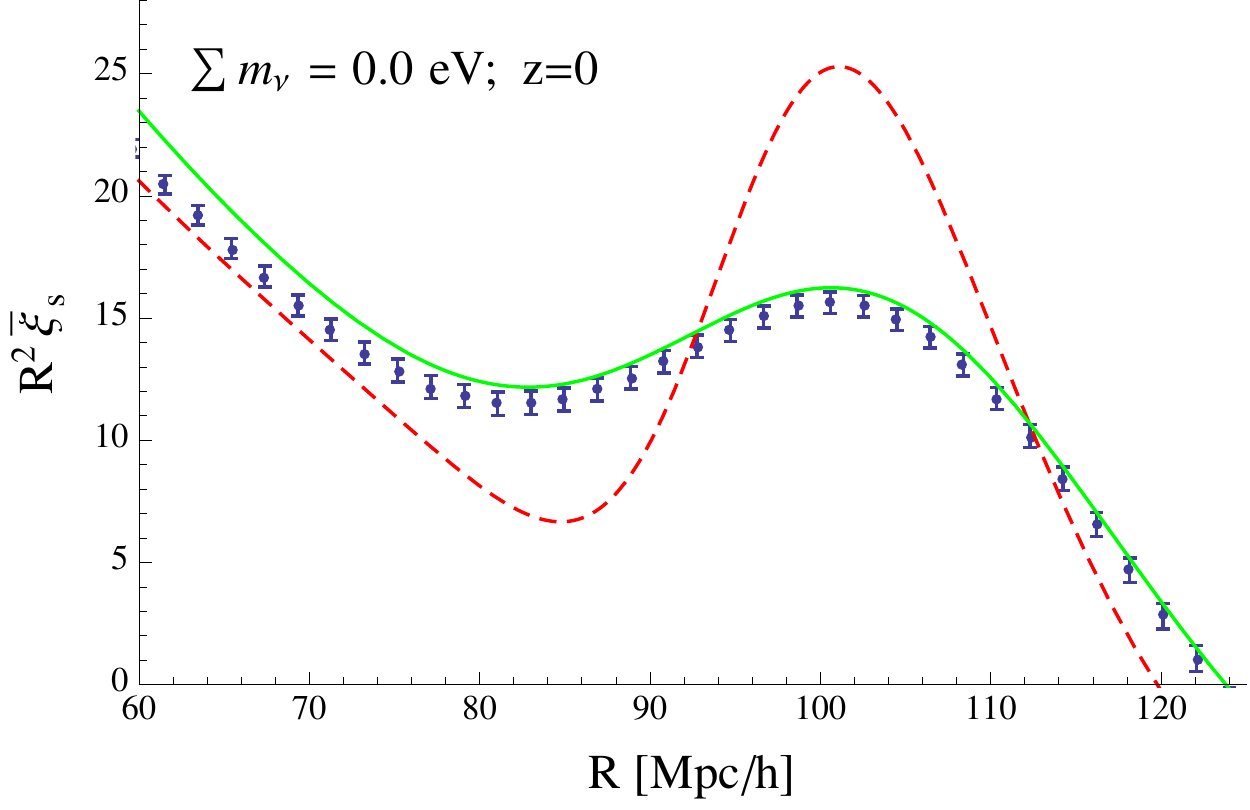}
\includegraphics[width=.45\textwidth,clip]{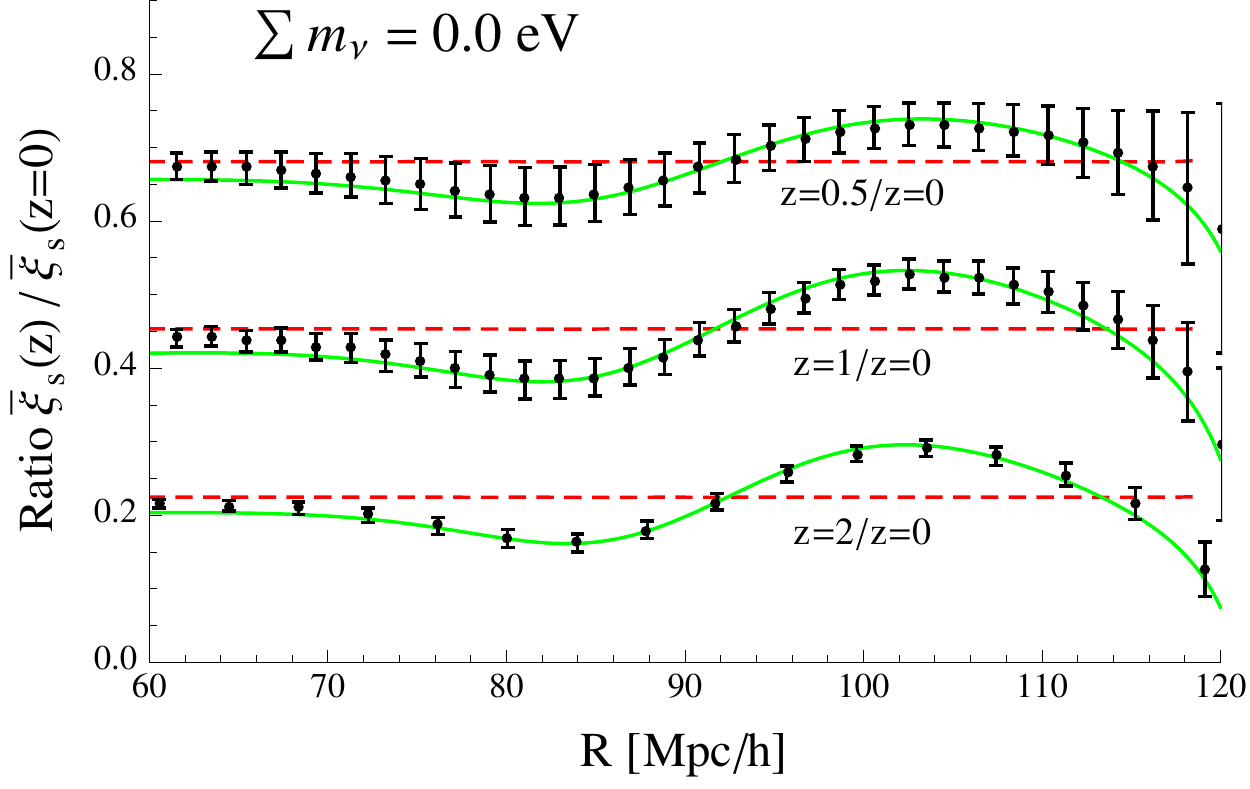}
}
\centering{ 
\includegraphics[width=.45\textwidth,clip]{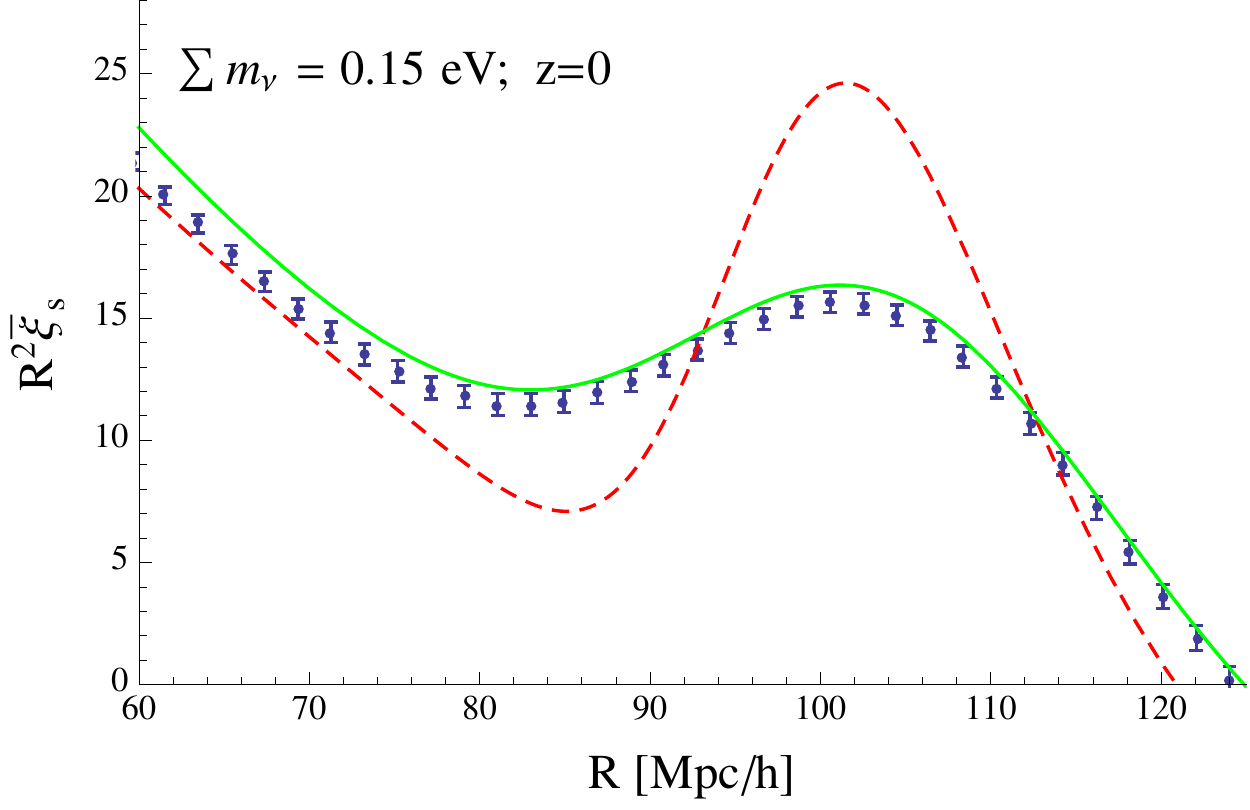}
\includegraphics[width=.45\textwidth,clip]{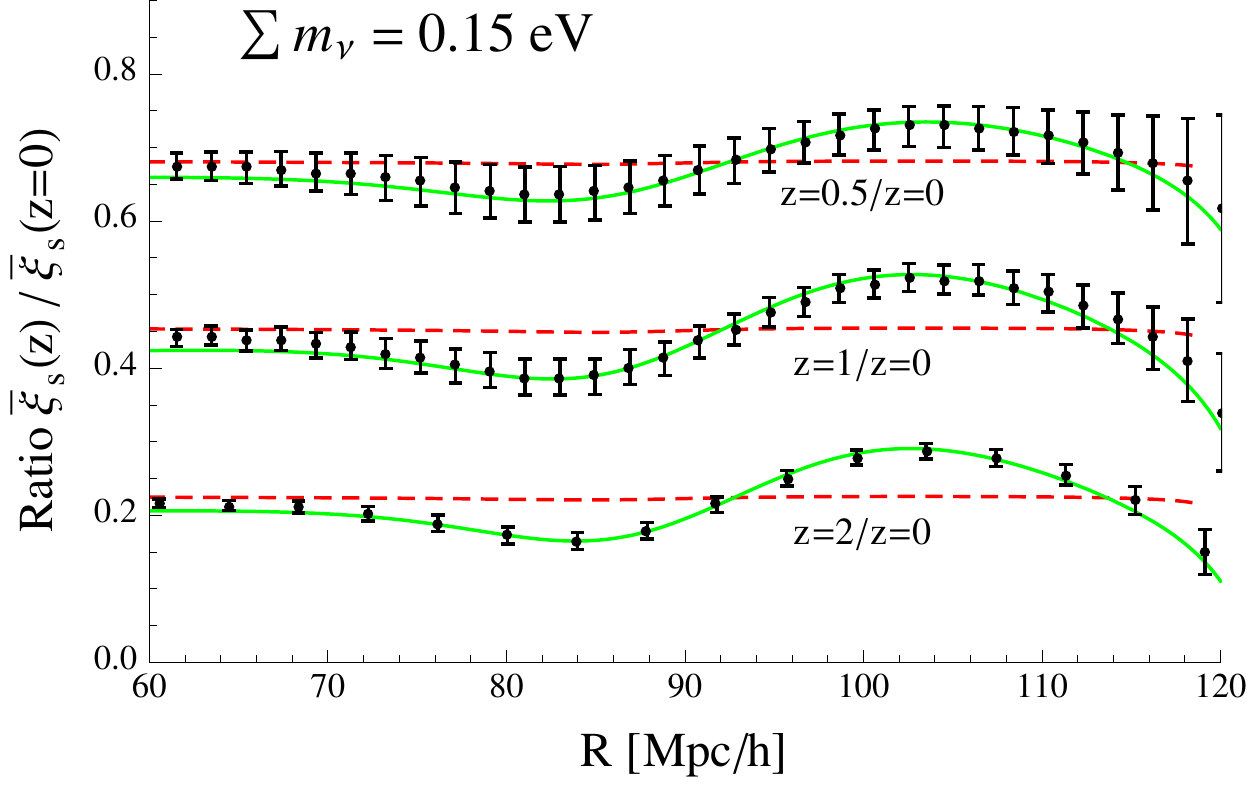}
}
\centering{ 
\includegraphics[width=.45\textwidth,clip]{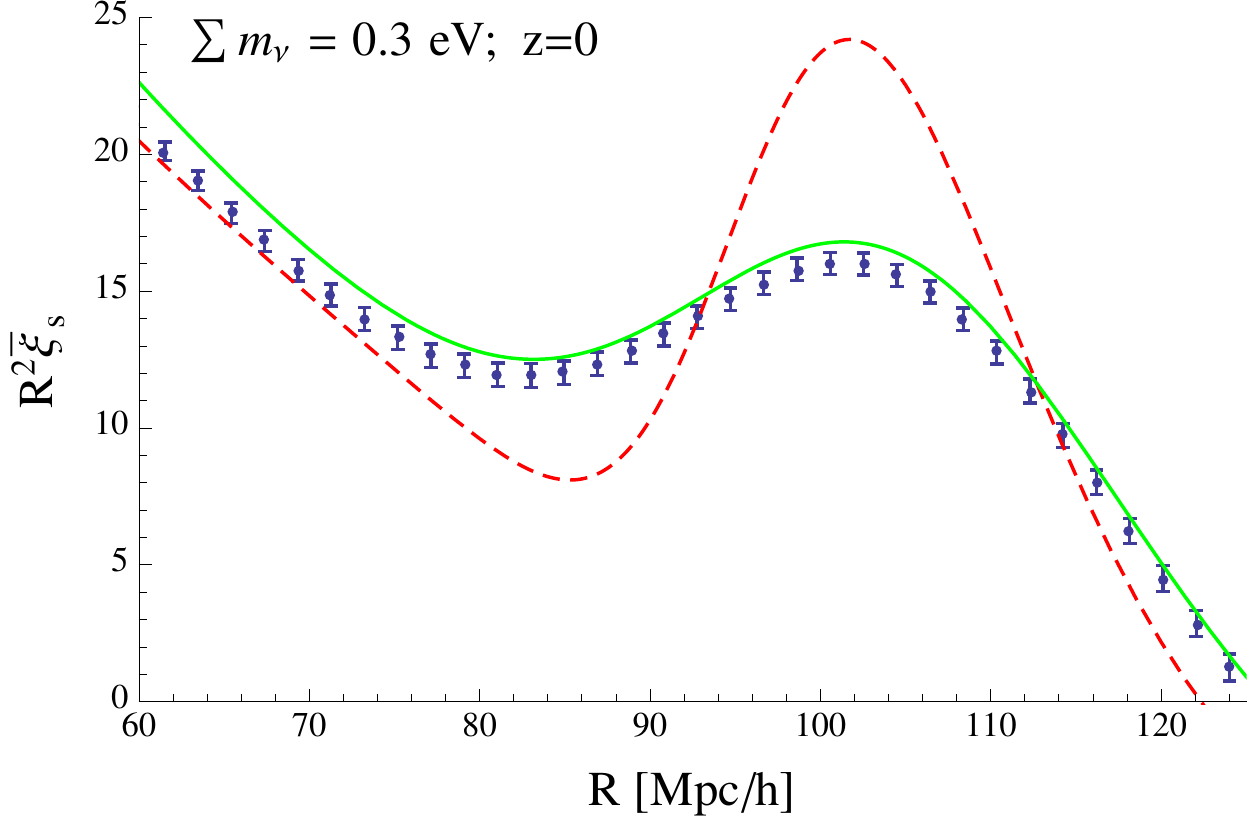}
\includegraphics[width=.45\textwidth,clip]{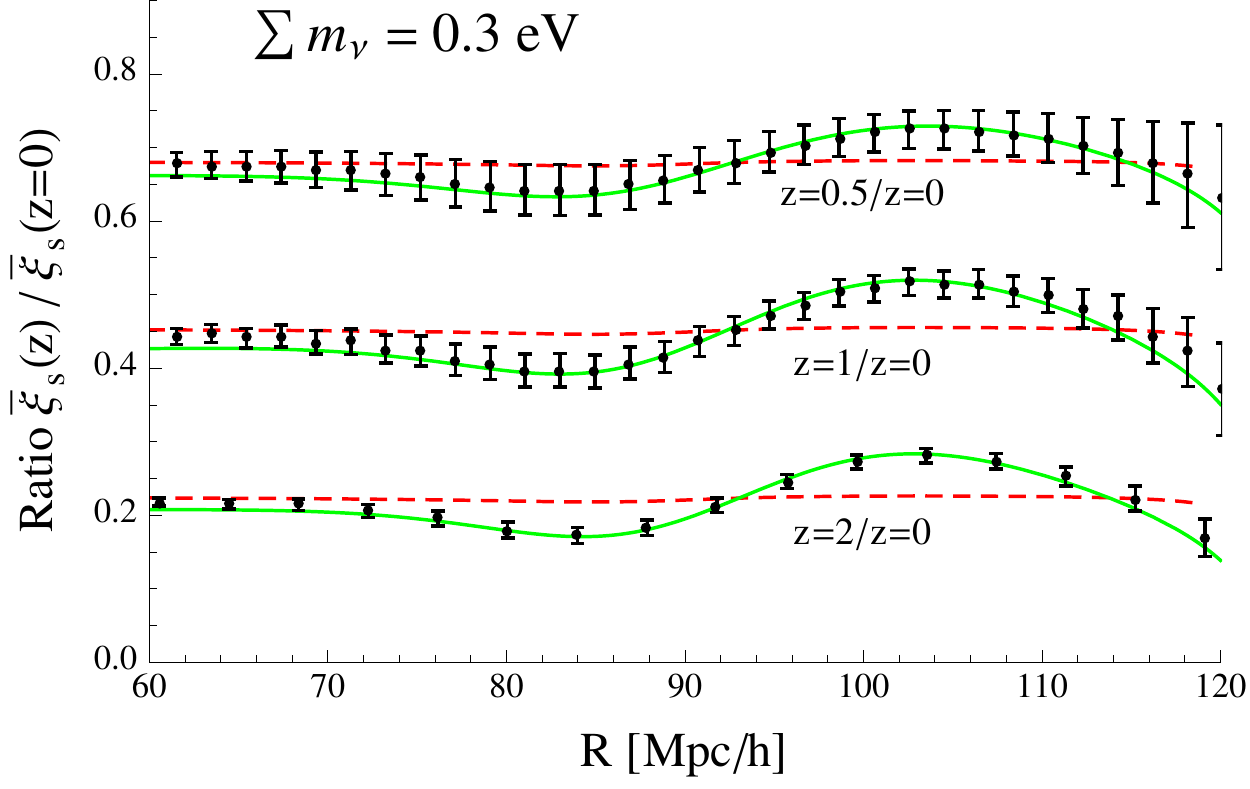}
}
\caption{\label{fig:CF-matter-redshift-m00-m015-m03} Matter CFs in redshift space at $z=0$ (first column) and ratios between CFs at different redshift (second column).  The figures in the first, second, third row are for $\sum m_\nu = 0 ,\, 0.15 ,\, {\rm and } \, 0.3$ eV, respectively. The data are from our N-body simulations;  the red dashed, and green solid lines are, respectively, for ${\bar \xi}_s^{\rm Kaiser}$ and for  ${\bar \xi}_s^{(1)}$,  defined in eq. (\ref{kaiser-1}).}
\end{figure}

\begin{figure}
\centering{ 
\includegraphics[width=.45\textwidth,clip]{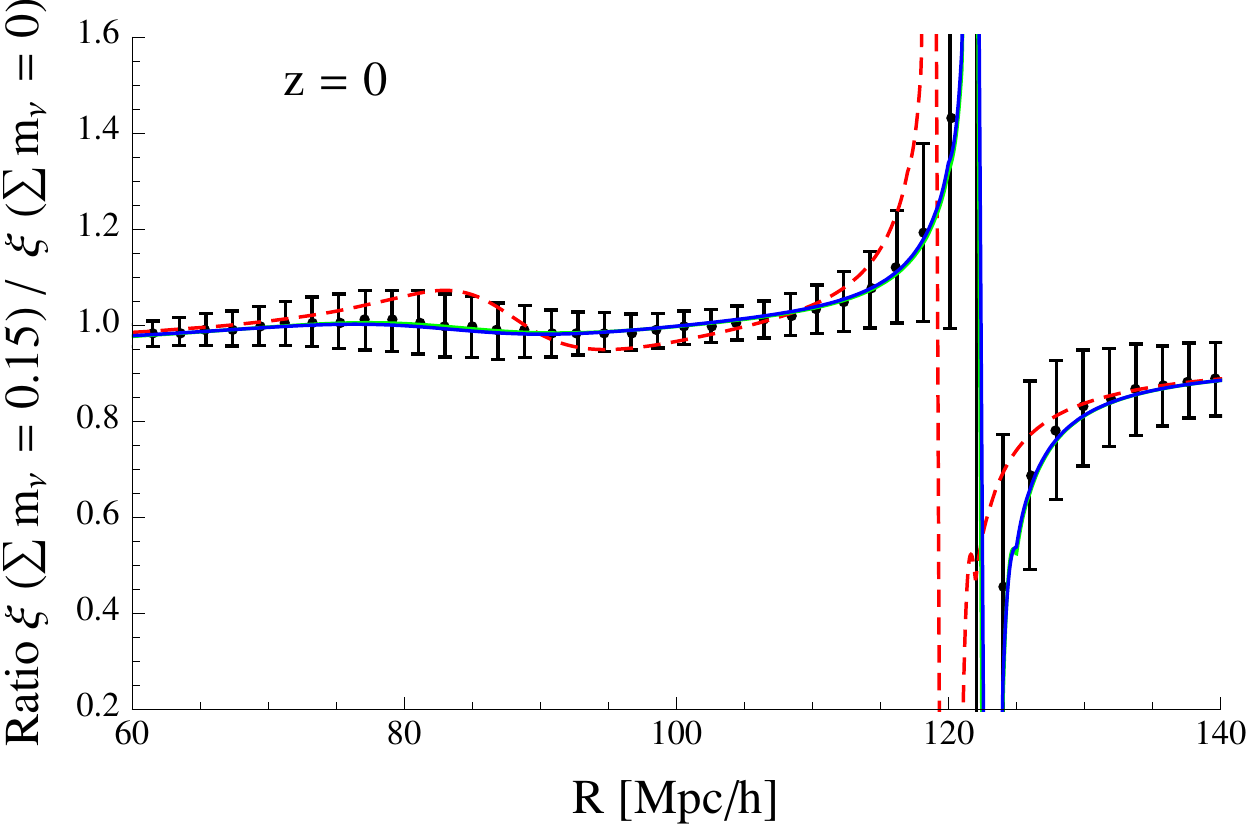}
\includegraphics[width=.45\textwidth,clip]{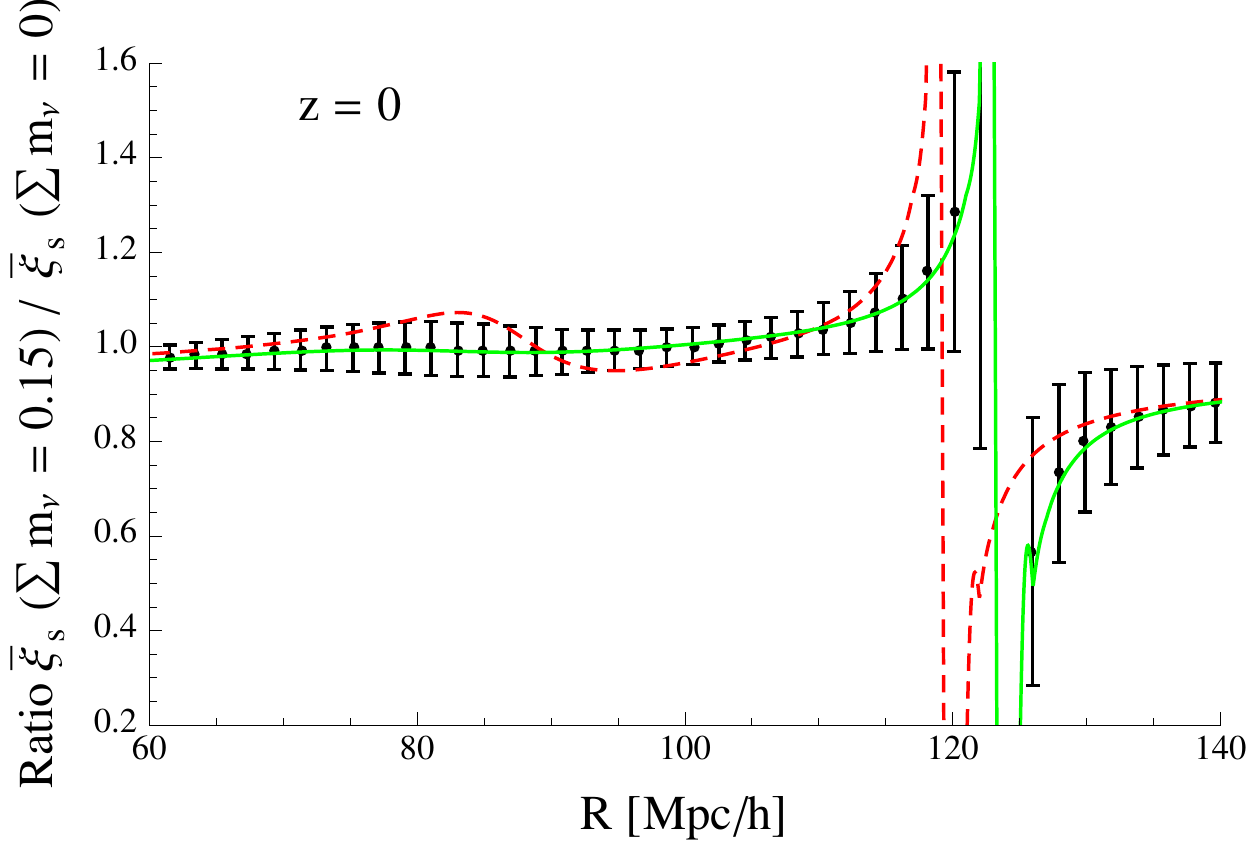}
}
\centering{ 
\includegraphics[width=.45\textwidth,clip]{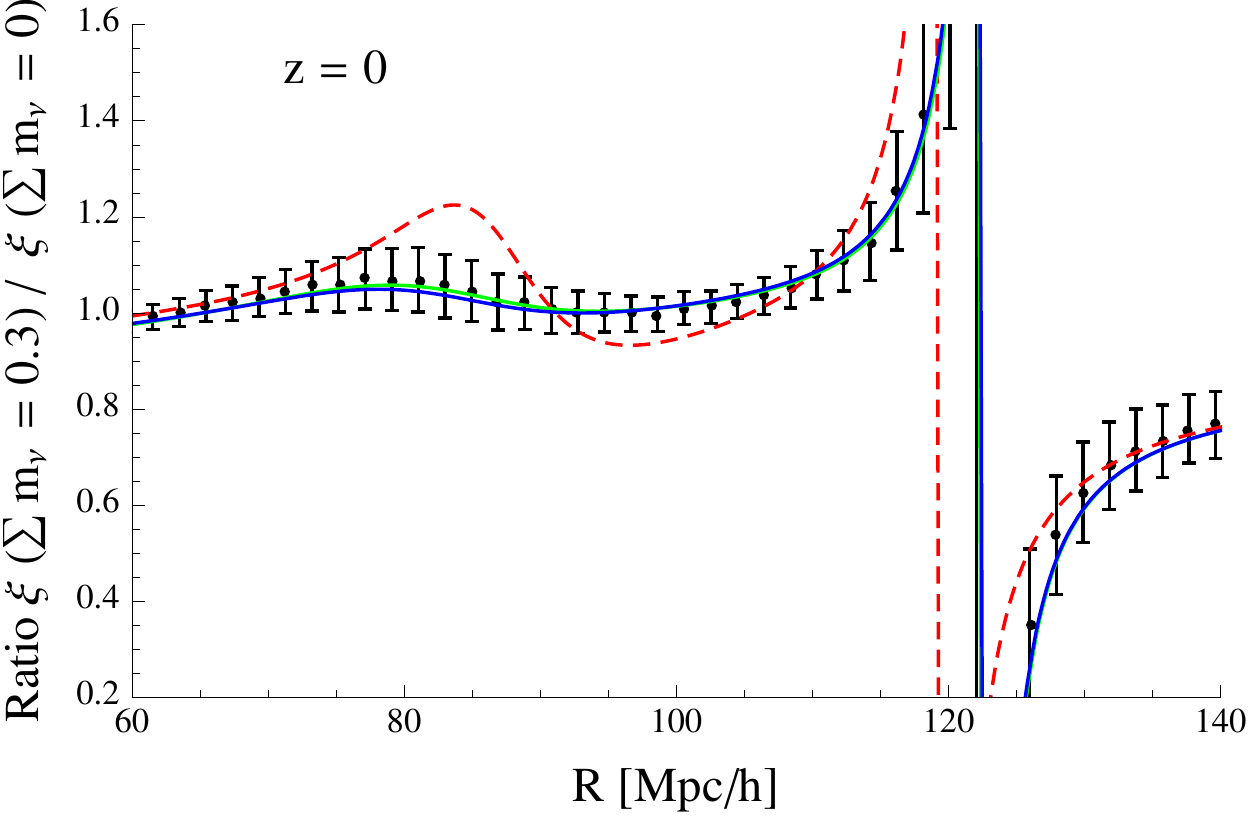}
\includegraphics[width=.45\textwidth,clip]{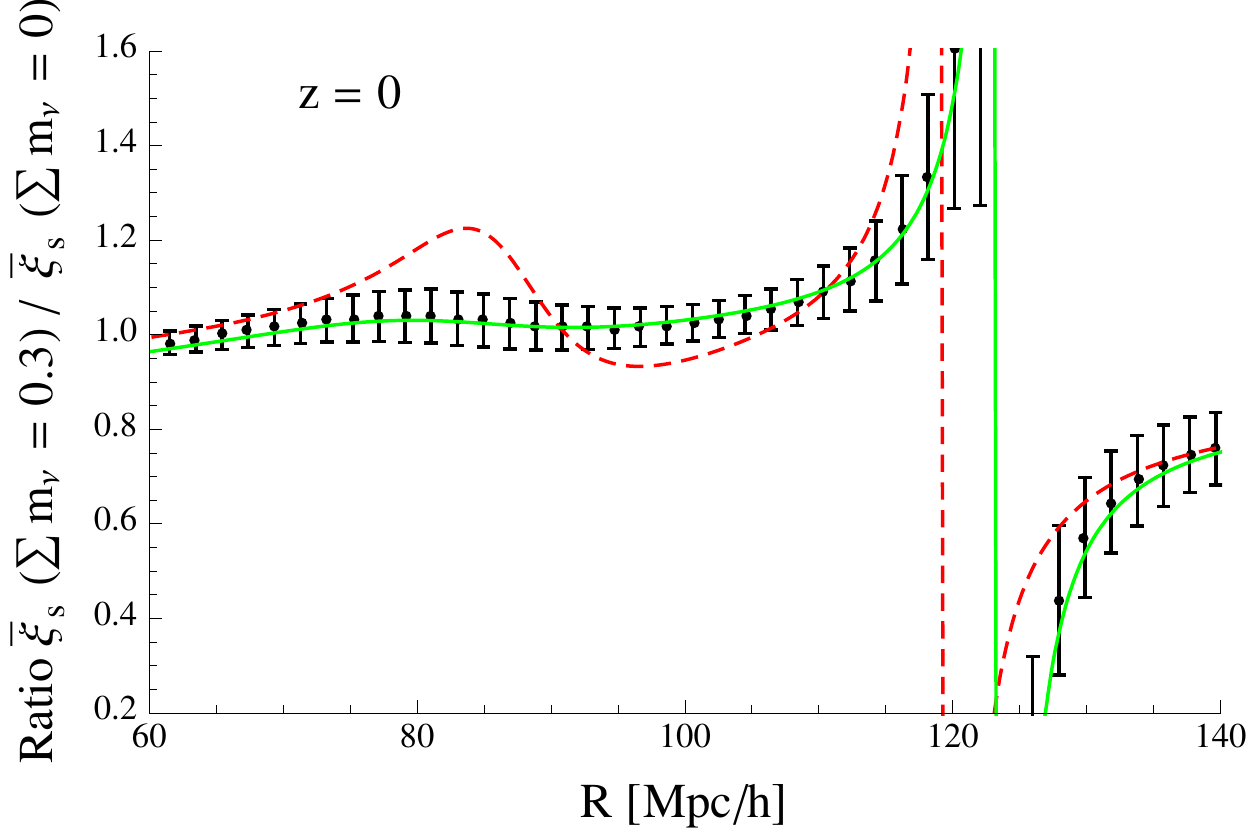}
}
\centering{ 
\includegraphics[width=.45\textwidth,clip]{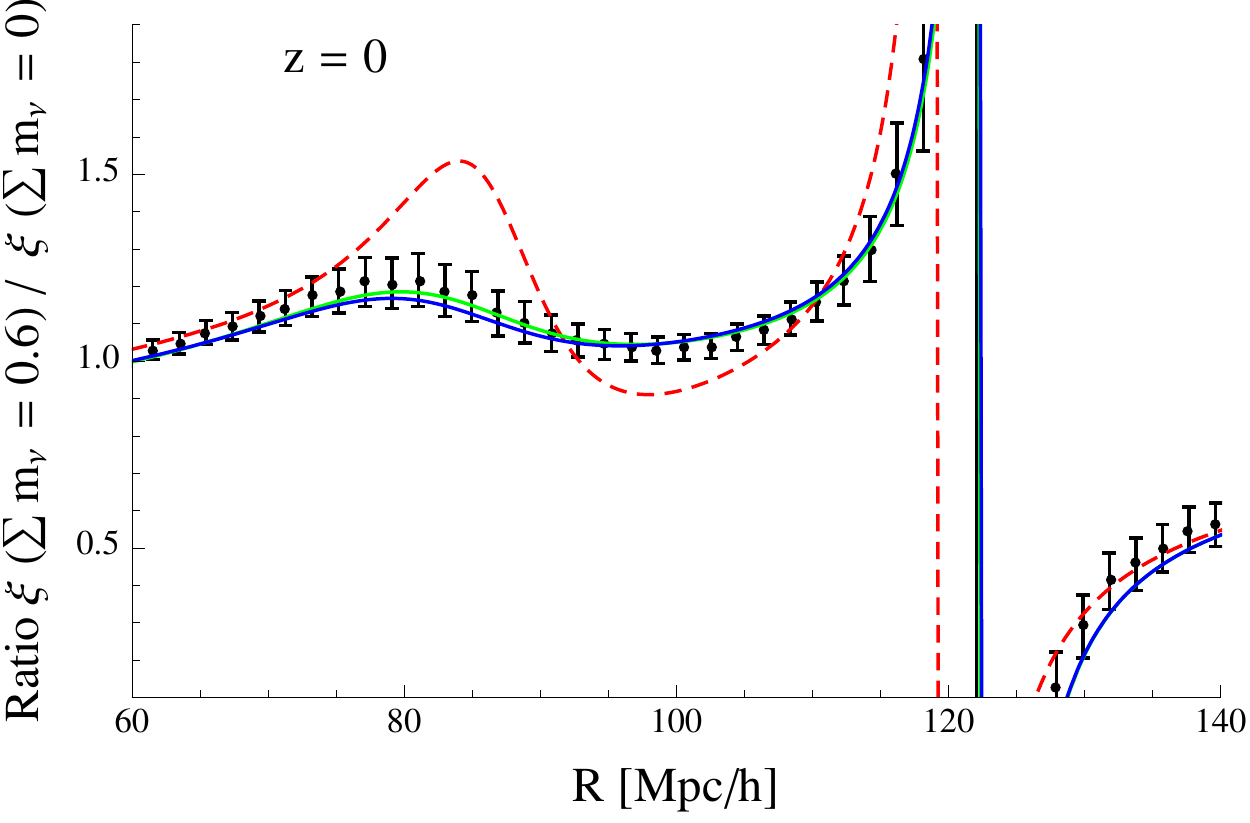}
\includegraphics[width=.45\textwidth,clip]{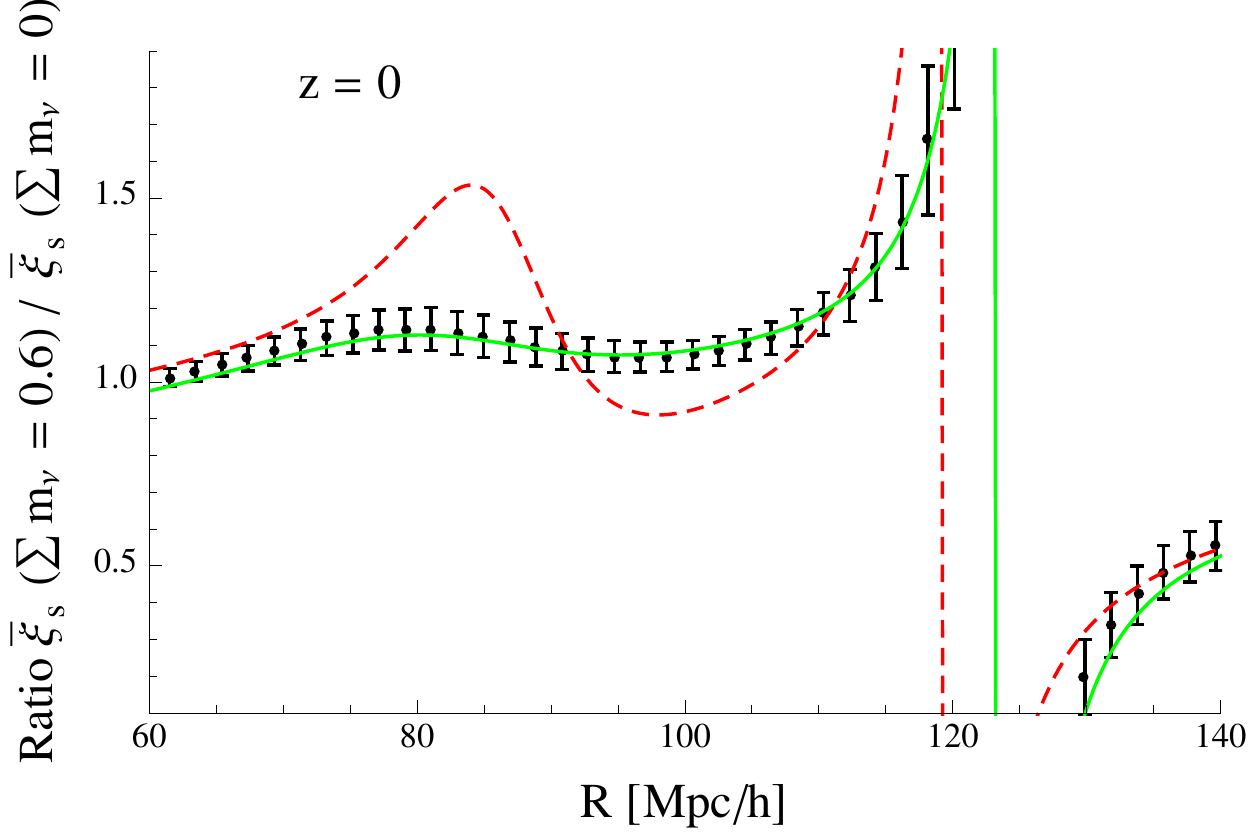}
}
\caption{\label{fig:CF-matter-real+redshift-ratm} Ratio between the $z=0$ matter redshift space CFs of two cosmologies with different neutrinos masses. The first, second, and third row show the CF for $\sum m_\nu = 0.15 ,\, 0.3 ,\, 0.6$ eV, respectively,  divided by the corresponding CF for massless neutrinos. The left column shows the ratios  in real space.  The data are ratios between our N-body simulations;  the red dashed, green solid, and blue solid lines are ratios between, respectively, $\xi^{\rm lin} ,\, \xi^{(1)}, {\rm and } \, \xi^{(2)}$,  defined in eq. (\ref{zeta-lin-1-2}). The right column shows the ratios in redshift space.  The red dashed, and green solid lines are, respectively,  ${\bar \xi}_s^{\rm Kaiser}$ and   ${\bar \xi}_s^{(1)}$,  defined in eq. (\ref{kaiser-1}).}
\end{figure}

\begin{figure}
\centering{ 
\includegraphics[width=.32\textwidth,clip]{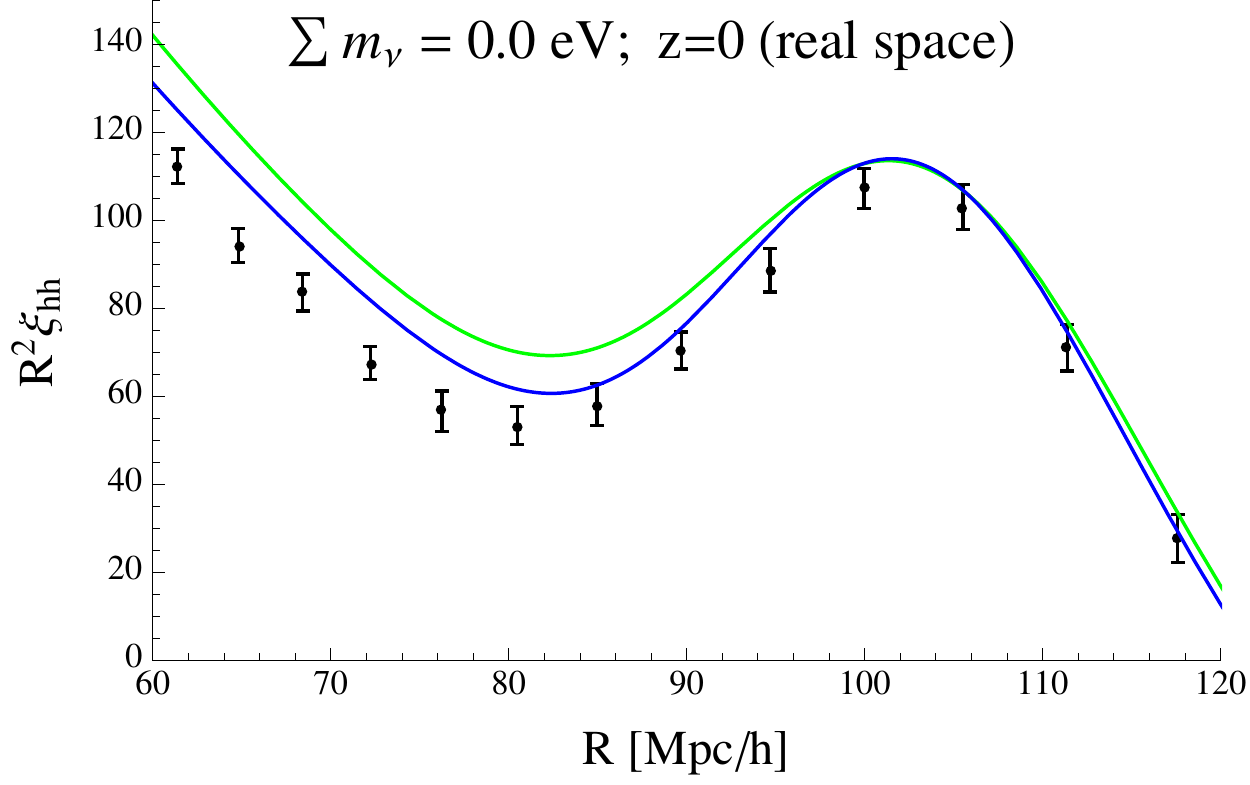}
\includegraphics[width=.32\textwidth,clip]{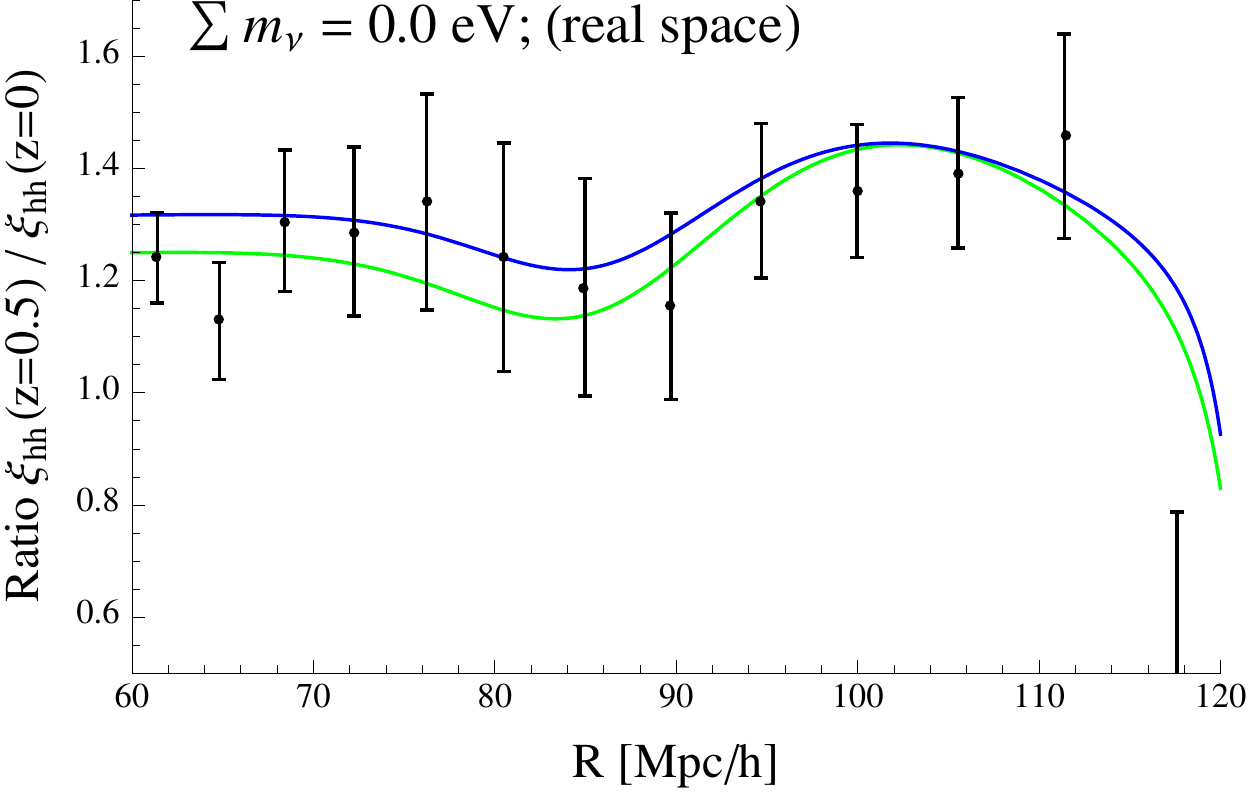}
\includegraphics[width=.32\textwidth,clip]{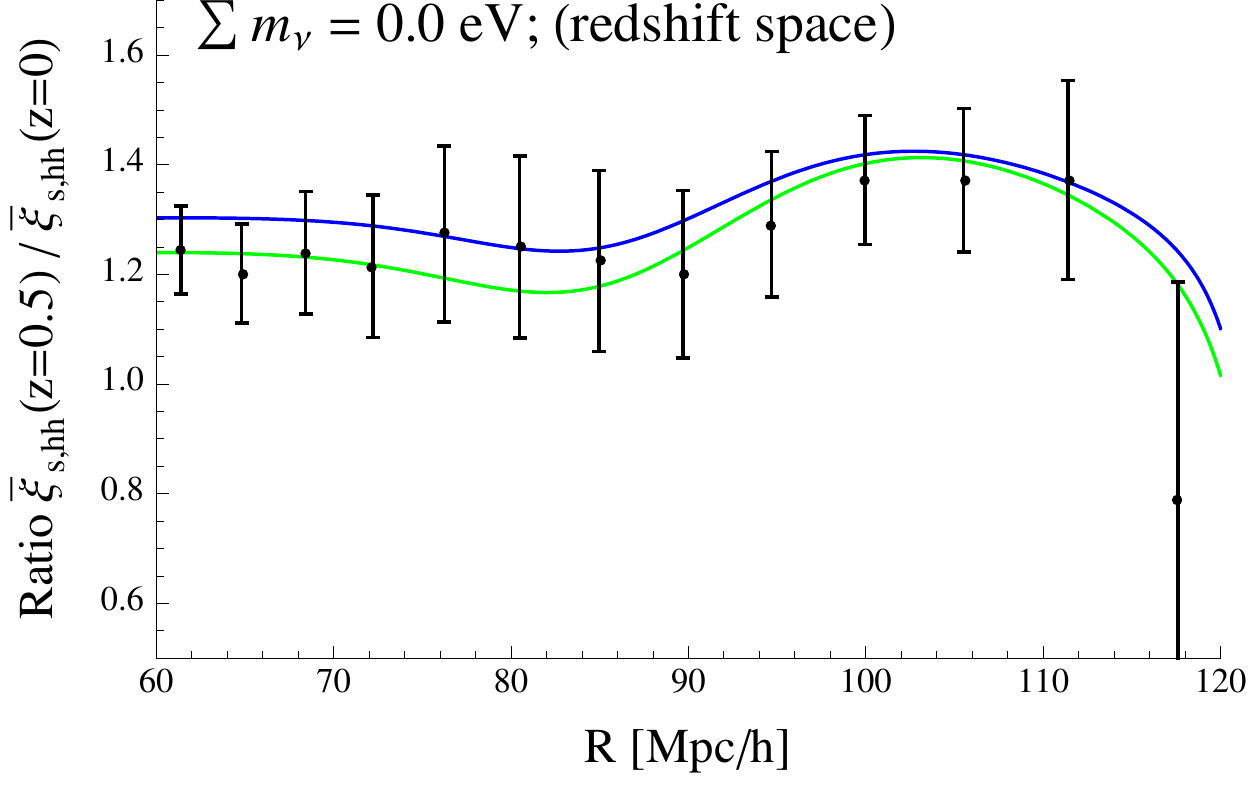}
}
\centering{ 
\includegraphics[width=.32\textwidth,clip]{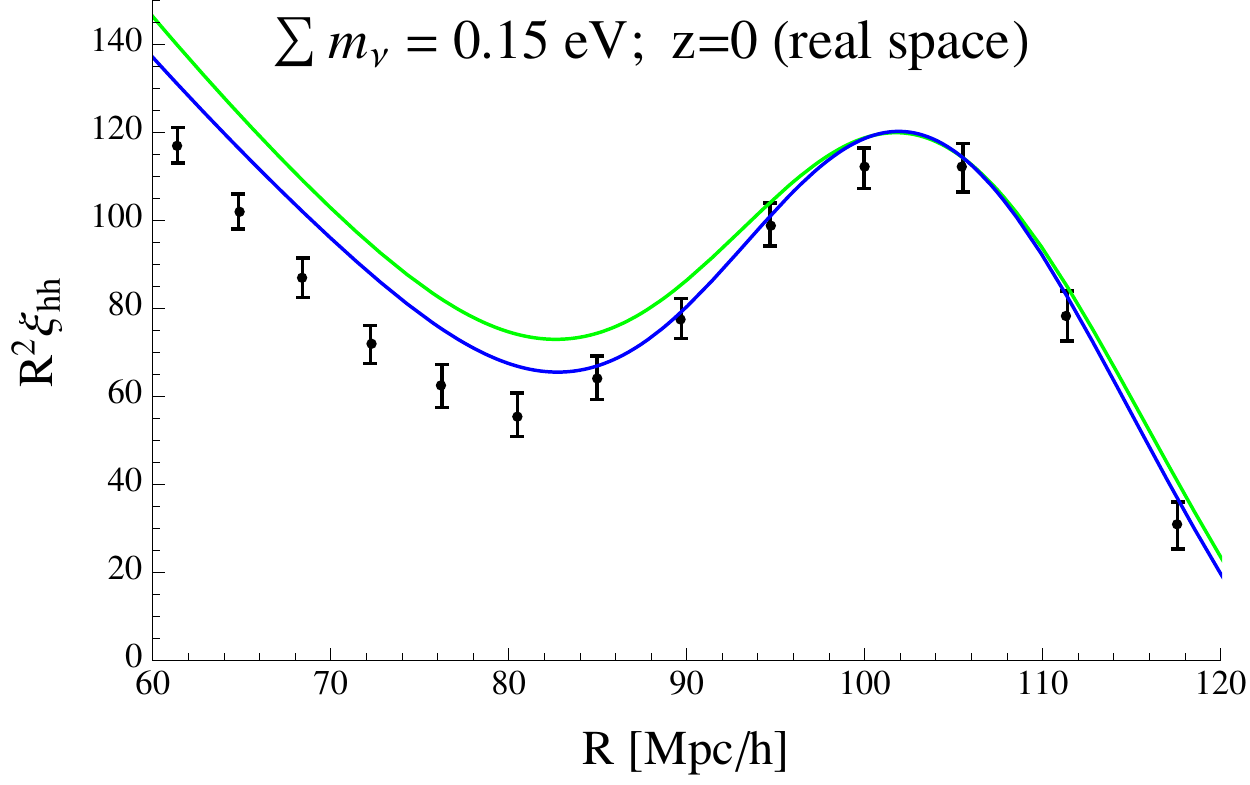}
\includegraphics[width=.32\textwidth,clip]{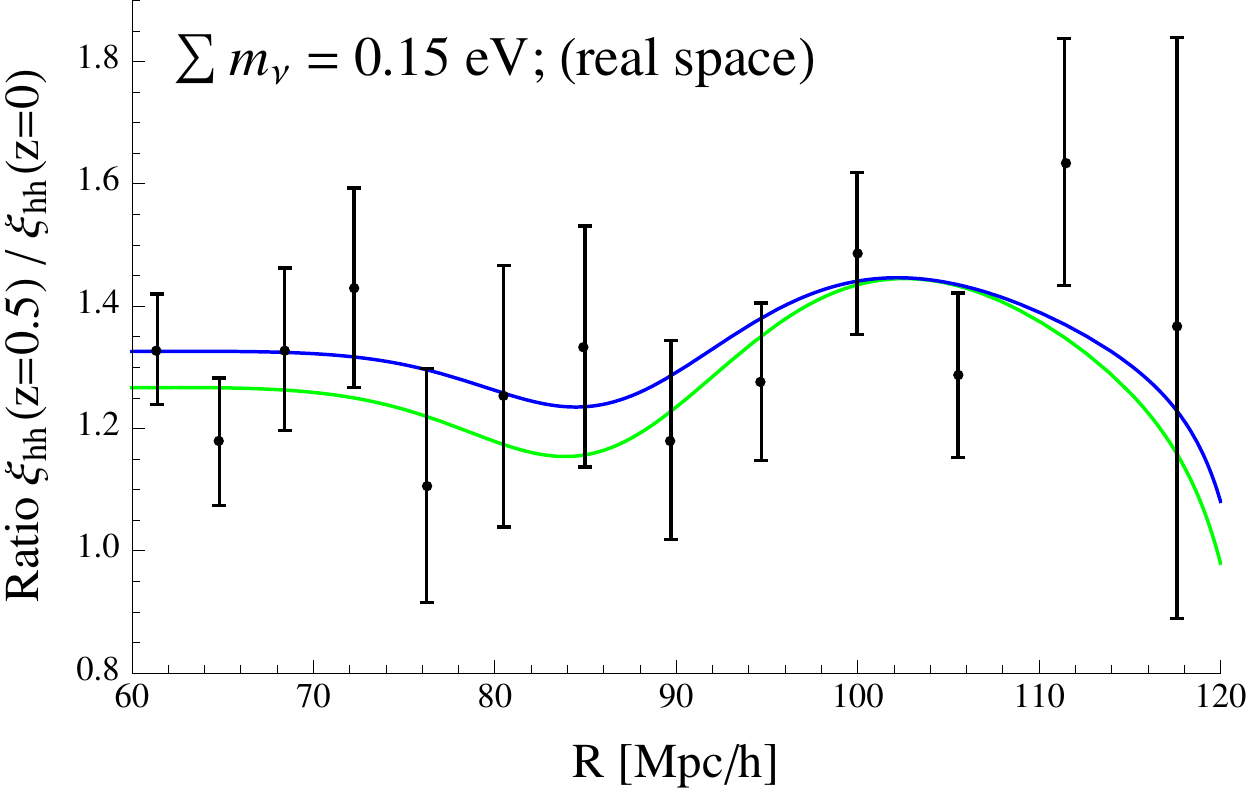}
\includegraphics[width=.32\textwidth,clip]{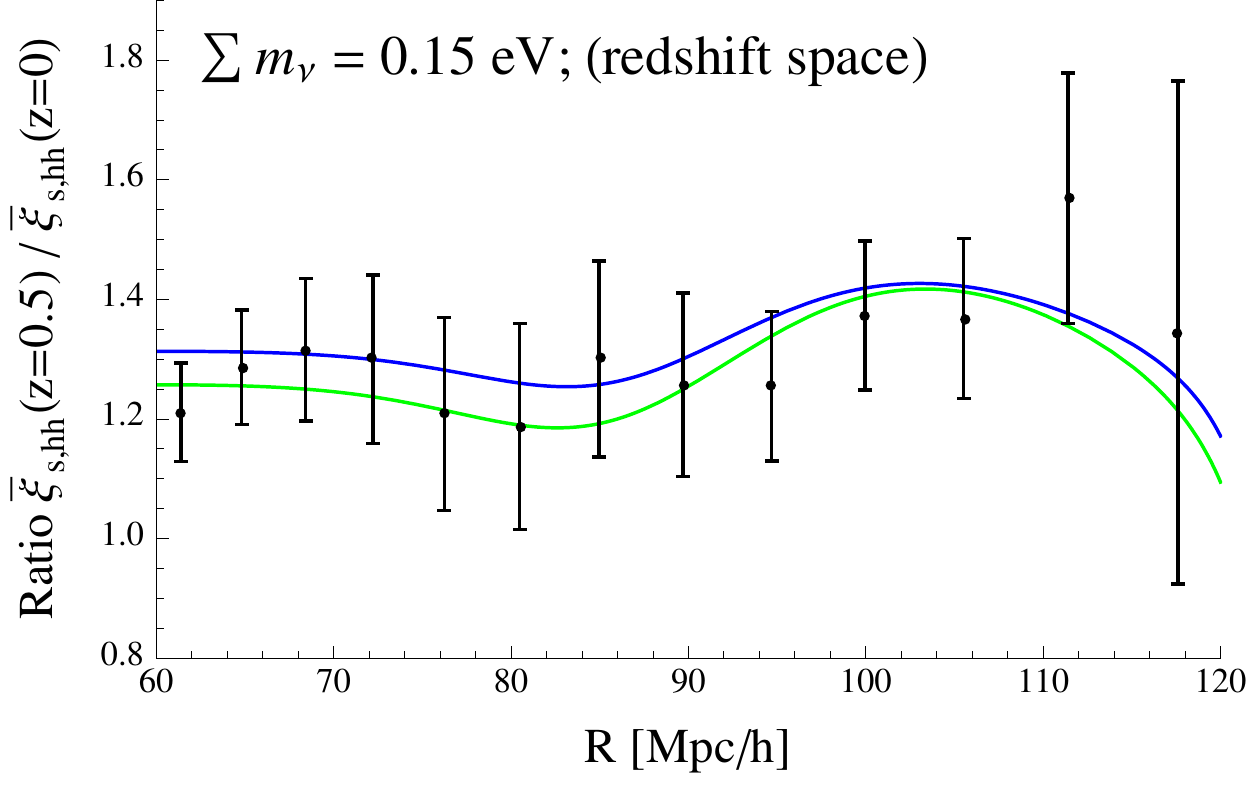}
}
\centering{ 
\includegraphics[width=.32\textwidth,clip]{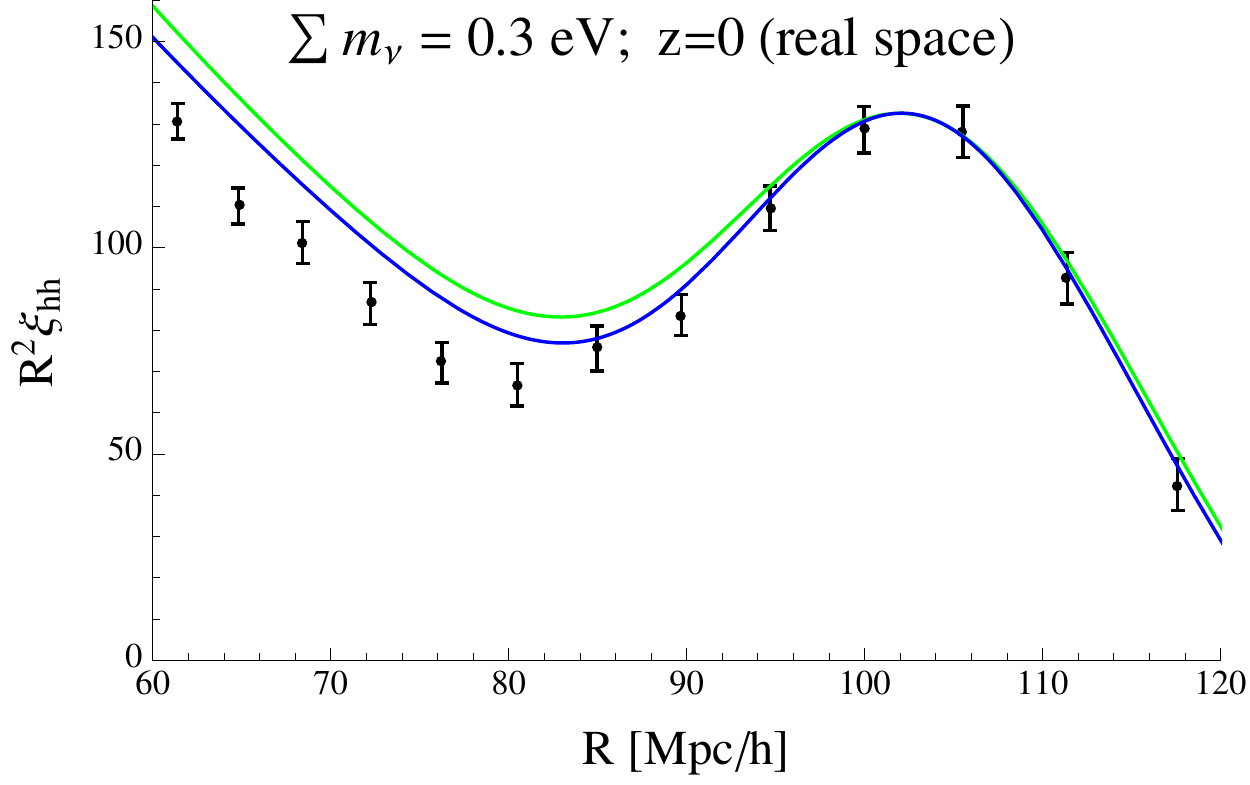}
\includegraphics[width=.32\textwidth,clip]{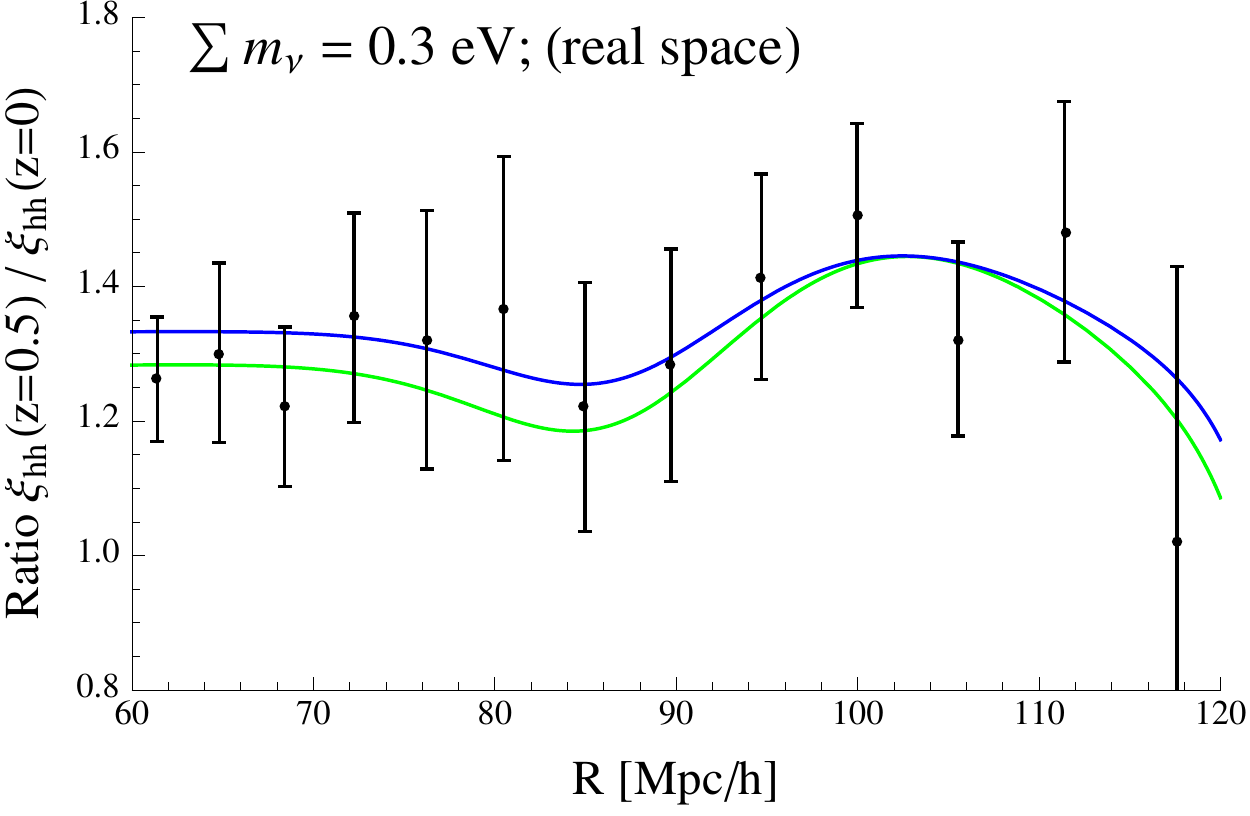}
\includegraphics[width=.32\textwidth,clip]{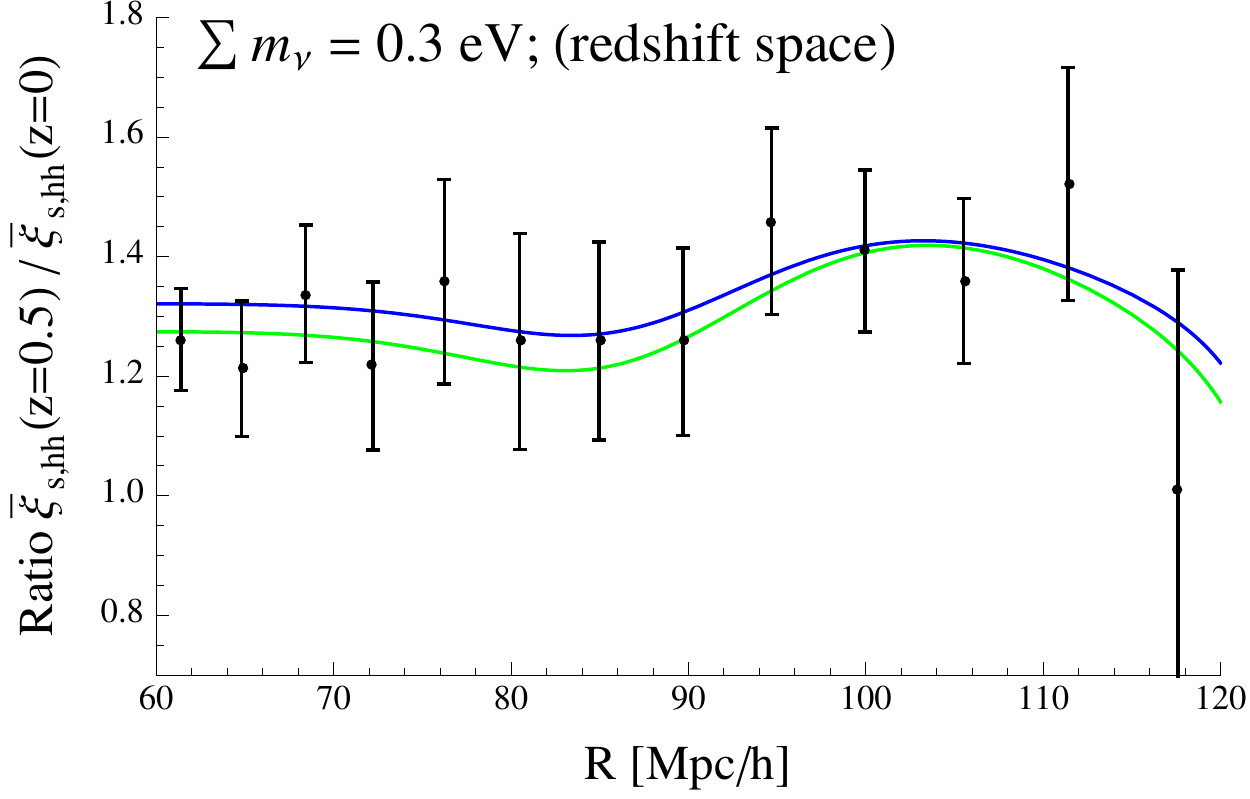}
}
\caption{\label{fig:CF-halo-z} Halo CFs. The figures in the first, second, and third row are for $\sum m_\nu = 0 ,\, 0.15 ,\, {\rm and } \, 0.3 $ eV, respectively. The left panels show the real space CF at $z=0$; the middle panels show the ratio between $z=0.5$ and $z=0$ CF in real space; the right panels show the same ratio in redshift space. The data are from our N-body simulations. The green (respectively, blue) curves are obtained from the bias function $b_{10}$ (respectively, $b_{10} +  b_{01} \, k^2$). 
}
\end{figure}

\begin{figure}
\centering{ 
\includegraphics[width=.45\textwidth,clip]{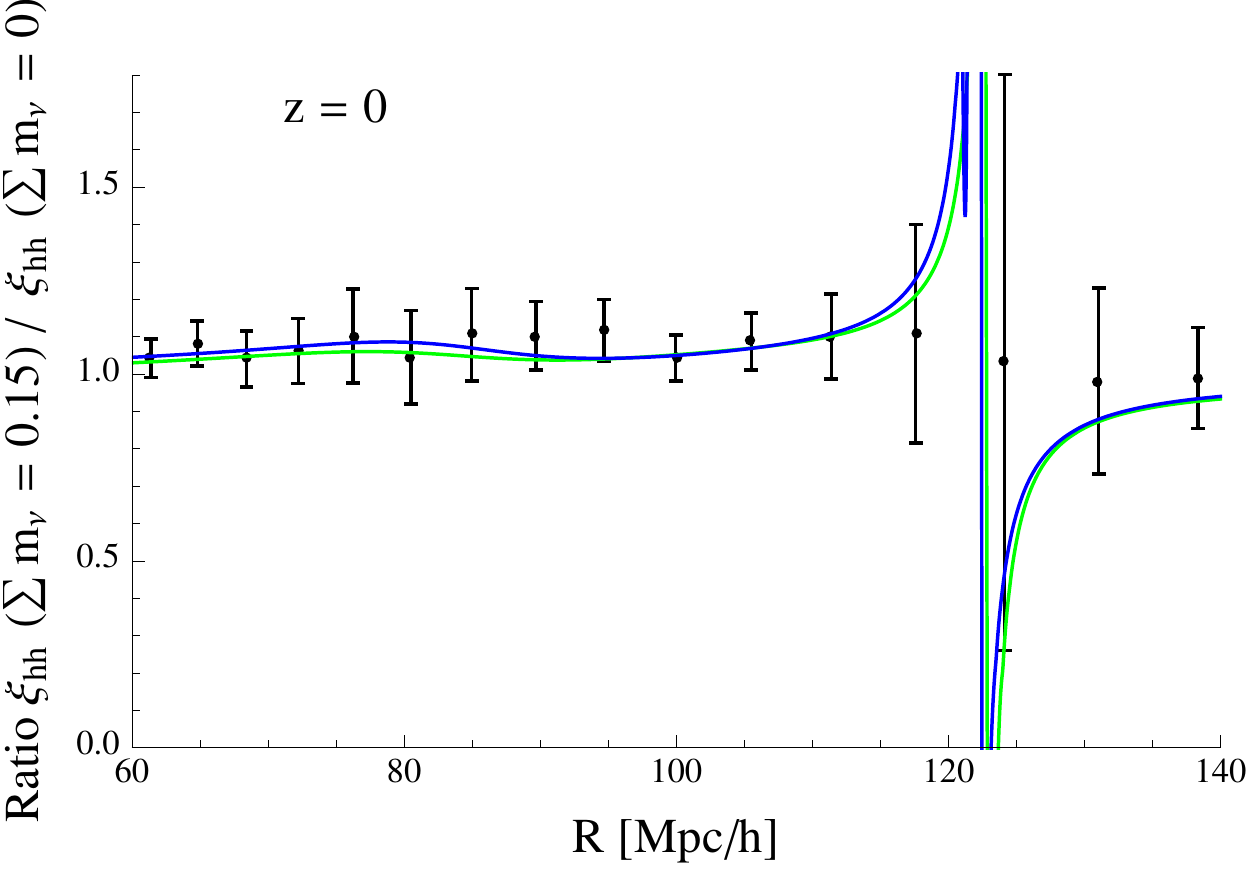}
\includegraphics[width=.45\textwidth,clip]{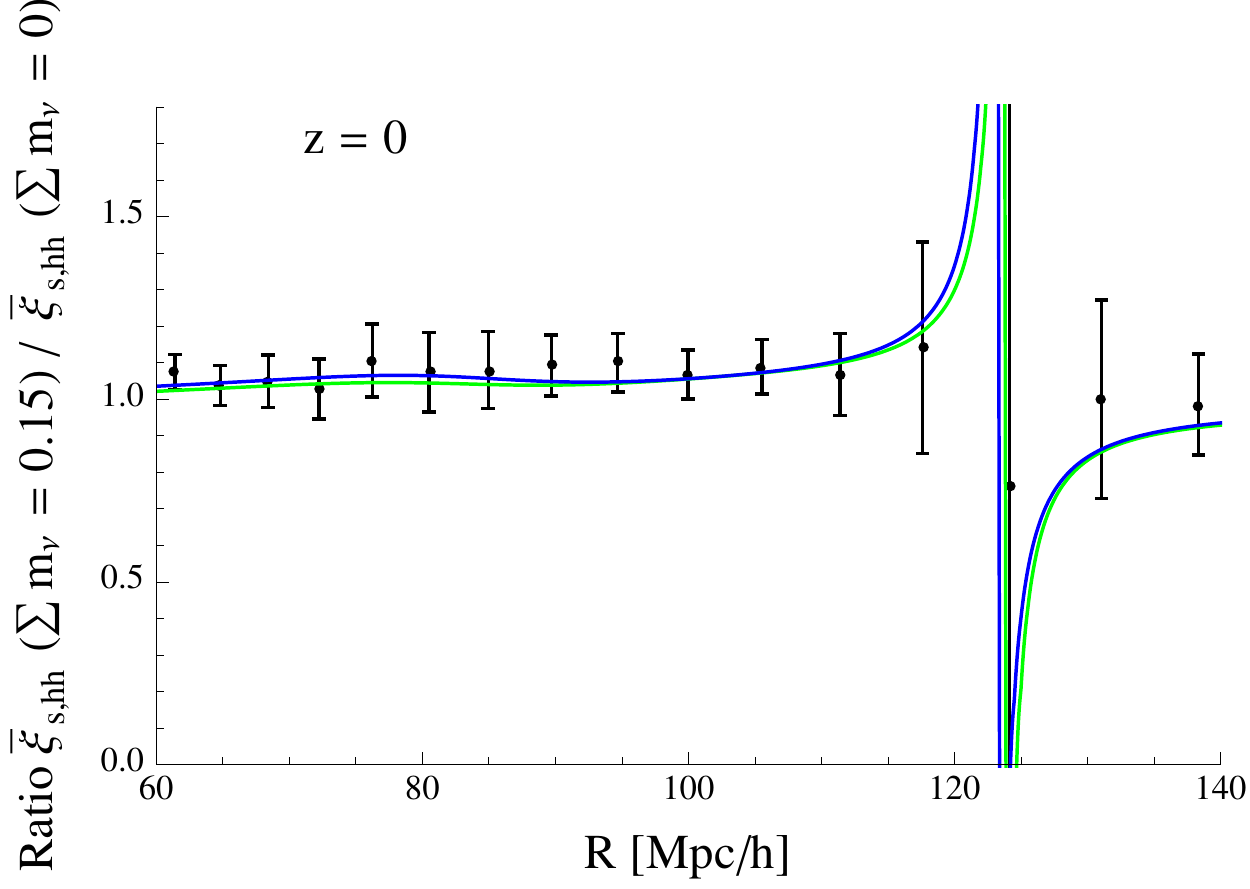}
}
\centering{ 
\includegraphics[width=.45\textwidth,clip]{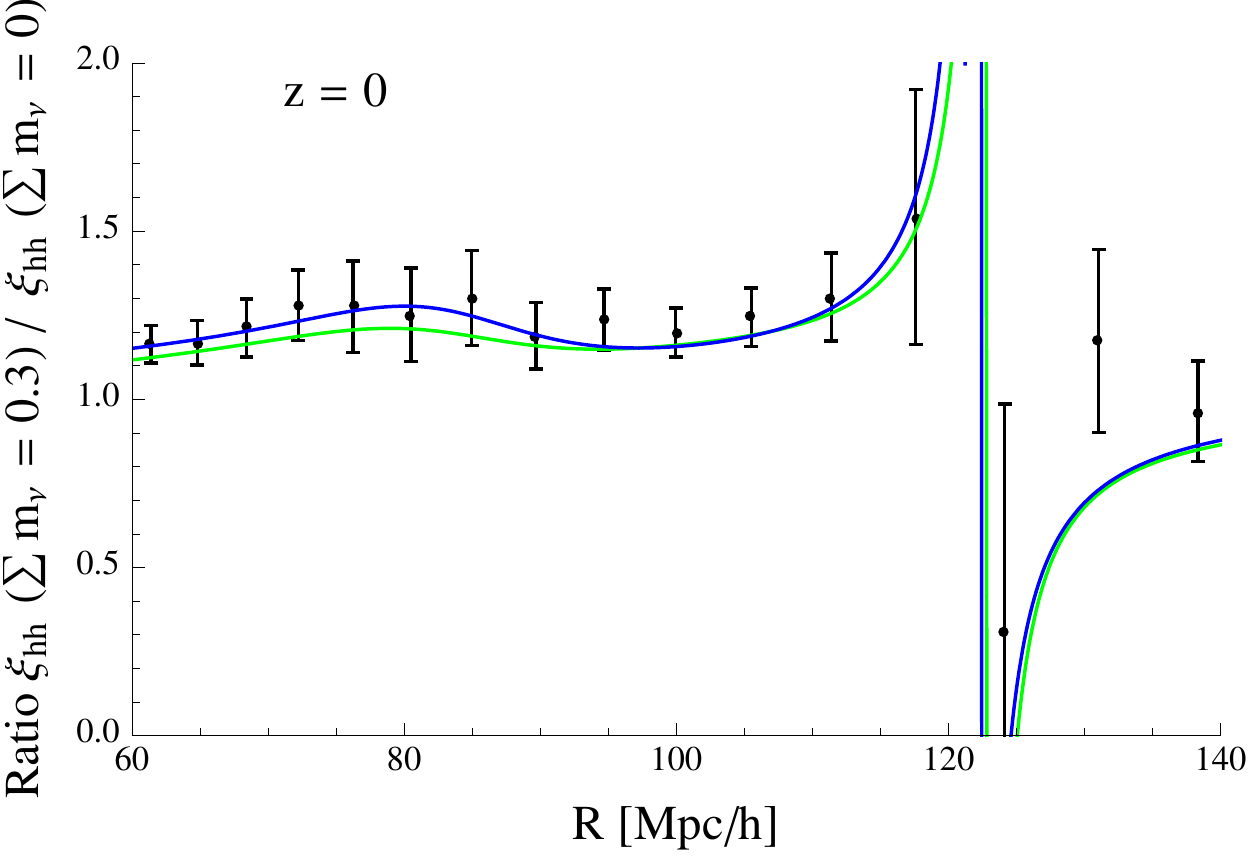}
\includegraphics[width=.45\textwidth,clip]{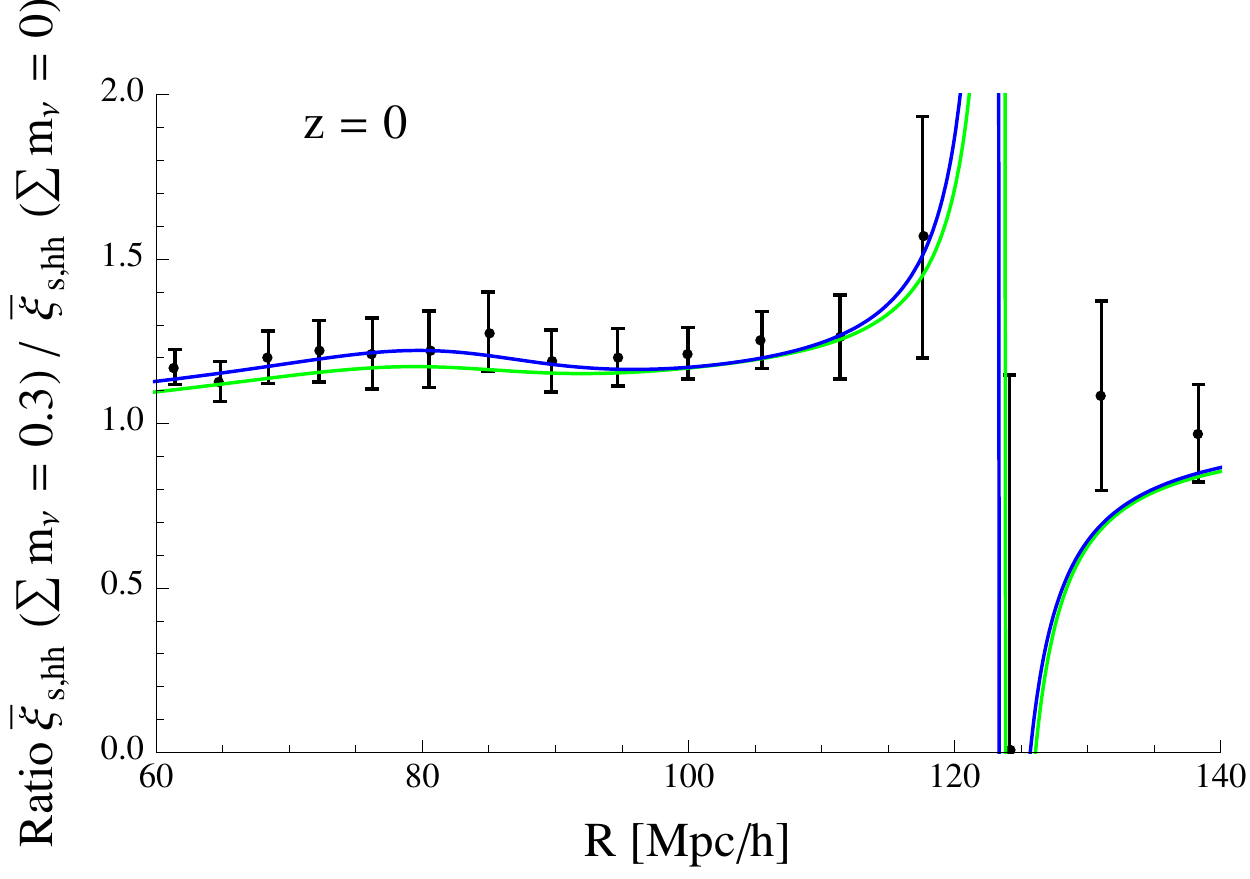}
}
\centering{ 
\includegraphics[width=.45\textwidth,clip]{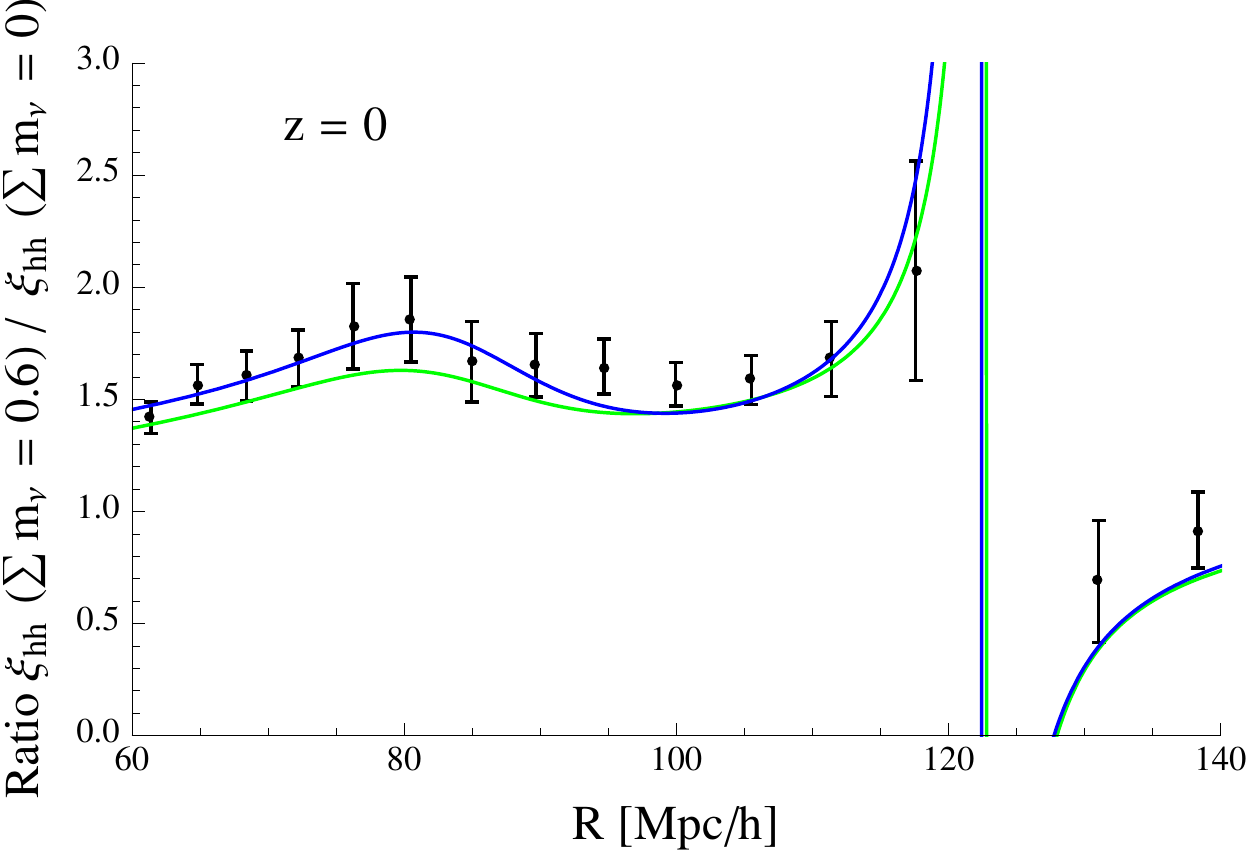}
\includegraphics[width=.45\textwidth,clip]{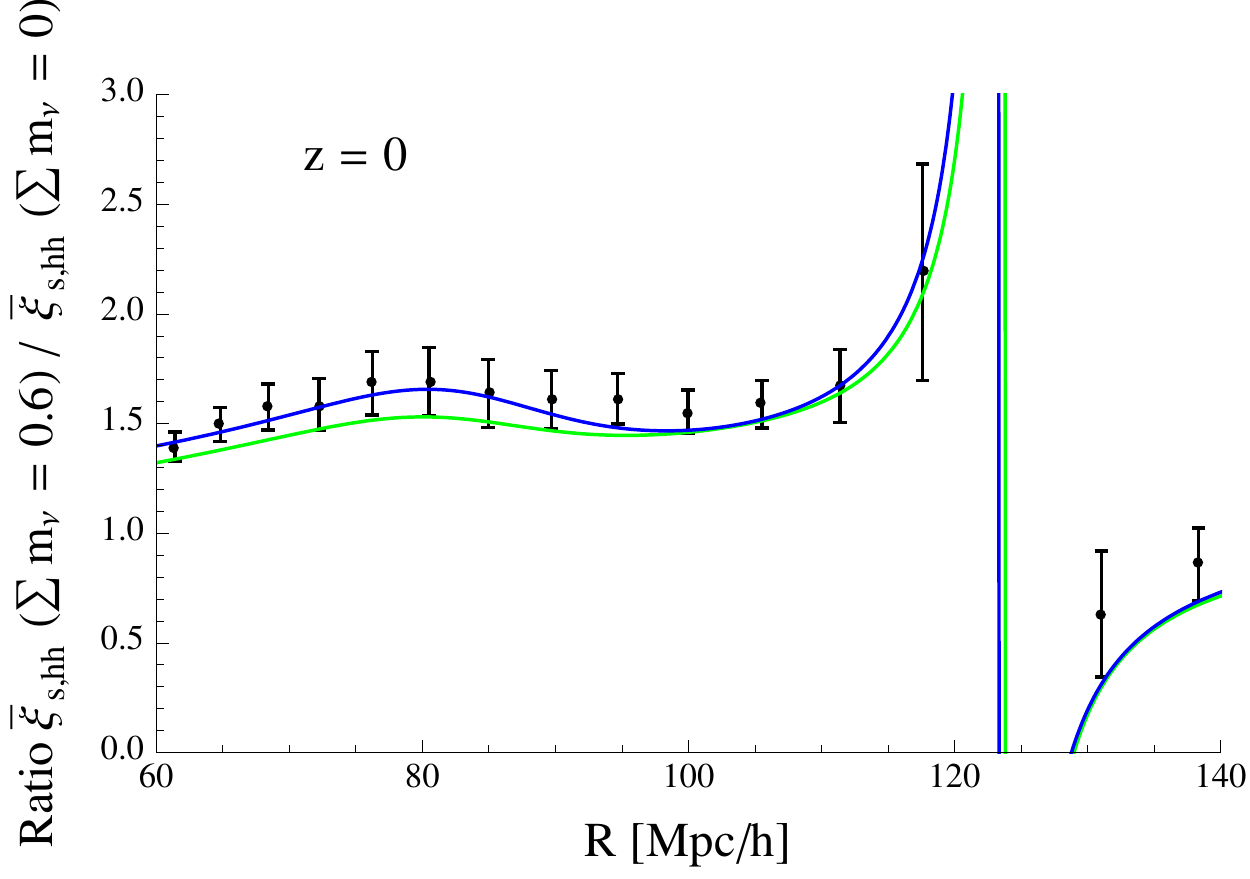}
}
\caption{\label{fig:CF-halo-m}  Ratio between the $z=0$ halo redshift space CFs of two cosmologies with different neutrinos masses.  The first, second, and third row show the CF for $\sum m_\nu = 0.15 ,\, 0.3 ,\, 0.6$ eV, respectively,  divided by the corresponding CF for massless neutrinos. The left (respectively, right) column shows the ratios  in real (respectively, redshift) space.  The data are from our N-body simulations. The green (respectively, blue) curves are obtained from the bias function $b_{10}$ (respectively, $b_{10} + k^2 \, b_{01}$). 
}
\end{figure}

We also note that the addition of the $1-$loop $P_{22}$ term in  $\xi^{(2)}$ shifts the peak to slightly smaller scales, at the subpercent level \cite{Smith:2007gi,RPTa, Seo:2009fp, Padmanabhan:2009yr, Sherwin:2012nh}, and it slightly improves the agreement with the FrankenEmu result at $z=0$ (the difference plotted as a black dashed line is both slightly closer to zero, and slightly more scale independent than the difference plotted as a black solid line; this is much less marked at the higher redshift shown, since the $1-$loop $P_{22}$ term becomes more irrelevant at increasing redshift). However, as we show in the next figures, the addition of the  $1-$loop $P_{22}$ term does not improve significantly the ratios between CF's.

Indeed, while the $\xi^{(1)}$ CF does not perfectly reproduce the N-body CF, it tracks extremely well how the CF changes with redshift. 
We see this from Figure \ref{fig:CF-matter-real-m00-zrat}, where we show ratios between matter CF (of the same cosmology) computed at different redshift. The ratios obtained from  $\xi^{(1)}$ are in excellent agreement with the ratios obtained from our N-body data, as well as with the FrankenEmu. We also see that, as we just mentioned, the inclusion of the $P_{22}$ term does not provide a significant improvement on these ratios.

Identical conclusions are obtained in the comparison between $\xi^{(1)}$ and our N-body data in the case of massive neutrinos. Notice that FrankenEmu does not provide data for these cosmologies. We show this in Figure \ref{fig:CF-matter-real-m015-m03}, where we present the CF at $z=0$, and the ratio between CFs at different redshift, in the case of $\sum m_\nu = 0.15$ eV (first row) and $0.3$ eV (second row). In these cases, we computed the velocity dispersion $\sigma_v^2$ using the linear PS for total matter in eq.~\re{sigma2}, that is for $\delta_m = \Omega_{\mathrm{c}} \delta_{\mathrm{c}}+\Omega_{\mathrm{b}} \delta_{\mathrm{b}}+\Omega_\nu \delta_{\nu}$, as it is the source of the Poisson equation.

It is natural to ask whether an equivalent agreement takes place also in redshift space. This is confirmed by Figure \ref{fig:CF-matter-redshift-m00-m015-m03}, where we show the comparison between the angular-averaged redshift space CF 
(\ref{kaiser-1}) and the one obtained from the N-body data. The linear correlations functions in real and redshift space are related to each other by the Kaiser relation (\ref{kaiser-1}). Not surprisingly, this also overpredicts the BAO peak. On the contrary, the CF  ${\bar \xi}_s^{(1)}$ shows an equal agreement with the N-body simulations as its real space counterpart $\xi^{(1)}$.

The real space CF $\xi^{(1)}$ and its redshift space counterpart  ${\bar \xi}_s^{(1)}$ are an optimal tool to study the dependence of the CF on the neutrino masses. This can be seen from Figure \ref{fig:CF-matter-real+redshift-ratm}, where we show ratios between the CF of a cosmology with massive neutrinos divided by a cosmology with massless neutrinos (the two cosmologies only differ from each other by the neutrino mass, and by the cold dark matter abundance, in such a way that $\Omega_{\rm m}$ is the same for them). Also in this case, the ratios obtained from  $\xi^{(1)}$ (left column plots) and   ${\bar \xi}_s^{(1)}$ (right column plots) are in excellent agreement  with the ratios obtained from the N-body data.

Actual measurements of the BAO peak involve biased objects. In Figure \ref{fig:CF-halo-z} we study the agreement between the halo CF $\xi_{hh}^{(1)}$ (\ref{CF1-hh}) and the halo CF obtained from our N-body simulations, as described in the previous section. The comparison is less probing than in the matter case, due to the increased sample variance of the latter (there are fewer halos than dark matter particles in the simulations). This is particularly true at increasing redshifts, and for this reason we only show halo data at $z=0,0.5$. The two solid lines shown in the figure are obtained with either a constant density bias,  $b \left( k \right) = b_{10}$ (green line) or a bias of the type  $b \left( k \right) = b_{10} +  b_{01} \, k^2$  (blue line),  times the exponential suppression due to the bulk flows, see eq. (\ref{CF1-hh}).  The bias coefficients are obtained by fitting the N-body correlation function (\ref{fit-bias}) at  large scales. For the plots on the left column we see that already using the constant bias allows to reproduce the height of the BAO peak. Allowing for the $b_{01} \, k^2$ term improves the agreement with the N-body CF at values of $R$ smaller than the peak. The plots in the  second and third column of the figure show ratios of CF at different redshifts (the second column shows ratios in real space, while the third column shows ratios in redshift space). We see that both the ratios obtained from a constant or from a linear bias are in agreement with the ratios from the N-body data (in fact, not appreciable scale dependence can be observed from the N-body ratios due to their error bars).

Finally, in Figure \ref{fig:CF-halo-m} we show ratios between the halo CF at $z=0$ of a cosmology with massive neutrinos and 
of a cosmology with massless neutrinos  (keeping the same $\Omega_m$ for all cosmologies, as we did for Figure \ref{fig:CF-matter-real+redshift-ratm}). We see that the ratios obtained using the $b_{10} + b_{01} \, k^2$ bias are in better agreement with the central value of the N-body data than those obtained from a constant bias, although also the latter appear to be within the $1-\sigma$ error bars of the  N-body data.

\section{Conclusions}
\label{conclusions}
The shape of the BAO peak contains cosmological information. In this paper we have focused on neutrino masses, and we have shown that, comparing cosmologies with the same $\Omega_{m}$, massive neutrinos with $ \sum m_\nu = 0.15$ eV   decrease the peak height by  $\sim - 0.6 \%$ with respect to the massless neutrino case, whereas  $\sum m_\nu = 0.3$ eV  (0.6 eV) increase it by $\sim 1.2\%$  ( $ 5.7 \%$).  These effects should be measurable in future surveys and can play an important role in breaking the known $\Omega_\nu$-$\sigma_8$ degeneracy from observables such as cluster number counts  \cite{Costanzi:2013bha}. 

The evolution of the BAO peak shape is dominated by the nonlinear process of random displacement of galaxies with respect to their initial locations. We have shown that it is reliably described by the propagator part of the nonlinear PS (the first term in \re{PSnl}), whereas the contribution from the mode-coupling part basically disappears when taking ratios of the CF computed at different redshifts, or for different values of the neutrino masses.  The propagator can be computed both in Lagrangian PT or in Eulerian PT after resummation (see, for instance, \cite{RPTb,MP07b,RPTBAO,Bernardeau:2008fa,Anselmi:2010fs,Anselmi:2012cn}), the leading result in both cases being the Zel'dovich propagator. The difference between the different schemes resides basically in the computation of the subleading corrections to the propagator and of the mode-coupling part, which, as we have seen, has moderate impact on the BAO peak shape. Therefore, any of the semi-analytic computation schemes proposed in the literature, based on resummations of the PT expansion in the Eulerian or in the Lagrangian frameworks, should give comparable performances as long as they give a mode-coupling independent part of the PS of the Zel'dovich form.

We have shown that the matter CF computed from the simplest formula for the PS, eq.~\re{PS1}, agrees with N-body simulations in real space. The effect of the same physical processes can also be taken into account in redshift space, leading to an agreement with simulations of the same quality as for the real space. 

In both cases, the quantity responsible for the BAO peak blurring by nonlinear effects is the velocity dispersion $\sigma_v^2$, computed in linear theory according to eq.~\re{sigma2}. In the massive neutrino cases we use the linear PS for the total matter density perturbation, consistently with the fact that this is the source of the gravitational potential in Poisson equation.

In the case of halos our results are consistent with no velocity bias (but this topic should be reconsidered with larger simulations) and a nonlocal density bias of the form  $b_{10} + b_{01} \, k^2$, as proposed in \cite{Desjacques:2008jj,Desjacques:2009kt,Baldauf:2014fza}.

The analysis performed in this paper for the neutrino masses can be extended to other cosmological parameters. In this regard, analytic formulae such as 
\re{shift} can be extremely useful for a fast assessment of parameter forecasting for future surveys.

\section*{Acknowledgments}
M. Peloso  acknowledges partial support from the DOE grant de-sc0011842  at the University of Minnesota.    M. Pietroni acknowledges partial support from the  European Union FP7  ITN INVISIBLES (Marie Curie Actions, PITN- GA-2011- 289442). M. Viel and F. Villaescusa-Navarro are supported by the ERC StG "cosmoIGM" and by INFN IS PD51 "INDARK" grants. Part of the results obtained in this work have been obtained using computing resources at the University of Minnesota Supercomputing Institute.
\appendix
\section{Structure of the nonlinear PS}
\label{nlPS}
In this appendix we derive the general expression \re{PSnl} for the nonlinear PS. We express the density and velocity divergence field, $\theta(\bk, z)\equiv i\, \bk \cdot \bv(\bk,z)$, using the compact notation \cite{RPTa} $\vp_a(\bk,z)$, where $\vp_1(\bk,z)=\delta(\bk,z)$ and $\vp_2(\bk,z)=-\theta(\bk,z)/f{\cal H}$.

We assume that the field $\vp_a(\bk,z)$ at a redshift $z$ is a deterministic functional of the initial density and velocity field\footnote{We assume here that vorticity and higher moments of the distribution function ({\it i.e.} velocity dispersion....) give negligible contribution to the scales we are interested in. However, the derivation can be easily extended to include initial conditions for these quantities as well, and the same form \re{PSnl} would be obtained in this case, after a proper redefinition of the propagator.  }, defined at some initial redshift $z_{in}$, $\vp^{in}_c(\bq)$, and therefore it can be expressed as a functional expansion
\beqra
&& \!\!\!\!\!\!\!  \!\!\!\!\!\!\! \vp_a[\vp^{in}](\bk,z) = \sum_{n=1}^{\infty}\frac{1}{n!}\int d^3 q_1\cdots d^3 q_n  \frac{\delta^n \vp_a[\vp^{in}](\bk,z)}{\delta\vp^{in}_{c_1}(\bq_1)\cdots \delta\vp^{in}_{c_n}(\bq_n) } \vp^{in}_{c_1}(\bq_1)\cdots \vp^{in}_{c_n}(\bq_n)  \,,\nonumber\\
\label{expa}
\eeqra
where momentum conservation implies that the coefficients of the expansion are $\propto \delta_D(\bk-\sum_{i=1}^n \bq_i)$.

The propagator is defined as the ensemble average of the first derivative of $\vp_a[\vp^{in}](\bk,z) $ with respect to $\vp^{in}_c$,
\beqra
&& \!\!\!\!\!\!\!  \!\!\!\!\!\!\!  \!\!\!\!\!\!\!  \!\!\!\!\!\!\!  \!\!\!\!\!\!\! \delta_D(\bk-\bq)\,G_{ac}(k,z)  \equiv \left \langle  \frac{\delta \vp_a[\vp^{in}](\bk,z)}{\delta \vp^{in}_c(\bq)} \right \rangle\,\nonumber\\
&& \!\!\!\!\!\!\!  \!\!\!\!\!\!\! \!\!\!\!\!\!\!  \!\!\!\!\!\!\! \!\!\!\!\!\!\!  =  \sum_{n=0}^{\infty} \frac{1}{n!}\int d^3 q_1\cdots d^3 q_n \, 
\frac{\delta^{n+1} \vp_a[\vp^{in}](\bk,z)}{\delta\vp^{in}_{c}(\bq)\delta\vp^{in}_{c_1}(\bq_1)\cdots \delta\vp^{in}_{c_n}(\bq_n) } \, \left  \langle \vp^{in}_{c_1}(\bq_1)\cdots \vp^{in}_{c_n}(\bq_n) \right \rangle\,, 
\label{propexp}
\eeqra 
where we have used translation  and rotation invariance of the ensemble averages, that is $ \langle \vp^{in}_{c_1}(\bq_1)\cdots \vp^{in}_{c_n}(\bq_n) \rangle\propto (2 \pi)^3 \delta_D(\sum_{i=1}^n \bq_i)$ and $G_{ac}(\bk,z)=G_{ac}(k,z) $. We note that we have not assumed that the initial conditions are gaussian.

The PS is given by
\beq
(2 \pi)^3 \delta_D(\bk+\bk^\prime) P_{ab}(k,z)  = \langle   \vp_a(\bk,z) \vp_b(\bk^\prime,z) \rangle\,.
\eeq
Expanding both functions as in \re{expa} we get a sum of terms containing field averages of the form 
\beq \langle  \vp^{in}_{c_1}(\bq_1)\cdots \vp^{in}_{c_n}(\bq_n)  \vp^{in}_{d_1}(\bp_1)\cdots \vp^{in}_{d_m}(\bp_m)\rangle\,,
\label{terms}
\eeq
where the first $n$ terms come from the expansion of $\vp_a(\bk,z) $ and the last $m$ ones from that of $\vp_b(\bk^\prime,z)$, and the sum over $n$ and $m$ is taken.
We can isolate all the terms in which only one of the $\vp^{in}_{c_i}(\bq_i)$ is averaged with one of the $\vp^{in}_{d_i}(\bp_j)$, and we will collectively indicate as ``mode coupling'' terms all the contributions to the above averages which are not of this form. Eq.~\re{terms} can then be rewritten as
\beqra
&& \!\!\!\!\!\!\!  \!\!\!\!\!\!\!  \!\!\!\!\!\!\!  \!\!\!\!\!\!\!  \!\!\!\!\!\!\! n \,m\; \langle \vp^{in}_{c_1}(\bq_1)\vp^{in}_{d_1}(\bp_1)\rangle 
 \langle  \vp^{in}_{c_2}(\bq_2)\cdots \vp^{in}_{c_n}(\bq_n)\rangle \langle   \vp^{in}_{d_2}(\bp_2)\cdots \vp^{in}_{d_m}(\bp_m)\rangle + {\mathrm{mode \;coupling}}\, \nonumber\\
&&  \!\!\!\!\!\!\!  \!\!\!\!\!\!\!  \!\!\!\!\!\!\!  \!\!\!\!\!\!\!  \!\!\!\!\!\!\! =  n \,m\; (2\pi)^3 \delta_D(\bq_1+\bp_1) P^{in}_{cd}(q_1)  \langle  \vp^{in}_{c_2}(\bq_2)\cdots \vp^{in}_{c_n}(\bq_n)\rangle \langle   \vp^{in}_{d_2}(\bp_2)\cdots \vp^{in}_{d_m}(\bp_m)\rangle \nonumber\\
&&+ {\mathrm{mode \;coupling}}\,,
\label{split}
\eeqra
where $(2 \pi)^3 \delta_D(\bq+\bp) P^{in}_{ab}(q)  = \langle   \vp_a^{in}(\bq) \vp_b^{in}(\bp) \rangle$. As we see in the ``mode-coupling" contributions the external momentum $\bk$ and $\bk^\prime$ are split in two or more momenta of the initial fields $\vp^{in}$. 
The $n\;m$ factor in front of the above expression comes from the symmetric integration and summation in \re{expa}. 

Using \re{split} and the expression for the propagator \re{propexp} we can write the nonlinear PS as
\beqra&&
\!\!\!\!\!\!\!\!\!\!\!\!  \!\!\!\!\!\! \!\!\!\!  \!\!\!\!\!\! \!\!\!\!  \!\!\! (2 \pi)^3 \delta_D(\bk+\bk^\prime) P_{ab}(k,z)  =\nonumber\\
&&\!\!\!\!  \!\!\!\!\!\! \!\!\!\!  \!\!\!\!\!\!  \!\!\!\!  \!\!\!\!\!\!  \int d^3 q\, d^3 p\;(2 \pi)^3  \delta_D(\bq+\bp)P_{cd}^{in}(q) \, \left \langle\frac{\delta \vp_a[\vp^{in}](\bk,z)}{\delta \vp^{in}_c(\bq)} \right \rangle \, \left \langle\frac{\delta \vp_b[\vp^{in}](\bk^\prime ,z)}{\delta \vp^{in}_d(\bp)} \right \rangle + {\mathrm{mode \;coupling}}\,,\nonumber\\
&& \!\!\!\!  \!\!\!\!\!\! \!\!\!\!  \!\!\!\!\!\!  \!\!\!\!  \!\!\!\!\!\!=(2 \pi)^3 \delta_D(\bk+\bk^\prime) \left( G_{ac}(k,z)G_{bd}(k,z) P_{cd}^{in}(k) +P^{MC}_{ab}(k,z) \right)\,. 
\eeqra

If we assume growing mode initial conditions then $\vp^{in}_1(\bq)=\vp^{in}_2(\bq)$, and therefore $P_{11}^{in}(q)=P_{12}^{in}(q)=P_{22}^{in}(q)=P^{lin}(q,z_{in})$. Moreover, if we are interested in the density PS we have to take $a=b=1$. This gives exactly the expression in \re{PSnl} with
\beq
G(k,z)\equiv G_{11}(k,z)+G_{12}(k,z)\,,\qquad\qquad P_{MC}(k,z)\equiv P^{MC}_{11}(k,z)\,.
\eeq

\section{Zel'dovich approximation}
\label{Zeldovich}
In this appendix we discuss the relation between formula \re{PS1} and the Zel'dovich approximation \cite{1995MNRAS.273..475S,RPTa,White:2014gfa}.
At a given time, which we omit in the following equations, particles have moved from the the initial (Lagrangian) positions $\bq$ by the dispacement field ${\bf \Psi}(\bq) = \bx-\bq$, where $\bx$ is the position of the particle which was in $\bq$ initially. The matter density perturbation, $\delta(\bk)$, can then be expressed as
\beq
\delta(\bk) = \int d^3 x\, \delta(\bx) e^{-i \bk\cdot\bx} = \int d^3 q\,e^{-i \bk\cdot\bq} \left(e^{-i \bk\cdot {\bf \Psi(\bq)}}-1\right)\,,
\eeq
by means of the mass conservation relation $\left(1+\delta(\bx)\right) d^3 x = d^3 q$.
Therefore, the PS is given by
\beq
P(k)=\int d^3 r \,e^{-i\bk\cdot\br} \left(\langle e^{-i\bk\cdot \Delta {\bf \Psi} }\rangle-1 \right)\,,
\eeq
where $\br=\bq-\bq'$ and $\Delta {\bf \Psi} \equiv  {\bf \Psi}(\bq)- {\bf \Psi}(\bq')$.
The Zel'dovich approximation now amounts to using the linear expression for the displacement field, ${\bf \Psi}(\bq) =i \int \frac{d^3 k}{(2 \pi)^3} \e^{i\bk\cdot\bq}\frac{\bk}{k^2} \delta^{lin}(\bk)$. Assuming $ \delta^{lin}(\bk)$ gaussian, the average of the exponent can be
evaluated as $\langle e^{-i\bk\cdot \Delta {\bf \Psi} }\rangle = e^{-1/2\langle( \bk\cdot   \Delta {\bf \Psi})^2\rangle}$, and we can write the PS as
\beq
P^{Zeld}(k)= \int d^3 r \,e^{-i\bk\cdot\br}\left( e^{-k^2 \sigma_{v}^2 } e^{I(\bk,\br)}-1\right)\,,
\label{Zeldf}
\eeq
where 
\beq
I(\bk,\br)=\int \frac{d^3 p}{(2 \pi)^3} \,\frac{\left(\bk\cdot\bp \right)^2}{p^4} \,\cos(\bp\cdot\br) P^{lin}(p)\,.
\eeq
Eq.~\re{Zeldf} provides the relation between the linear and nonlinear PS in the Zel'dovich approximation. It can be evaluated numerically and compared to simulations, see for instance \cite{White:2014gfa}. 
Expanding the second exponential in parenthesis in \re{Zeldf} we can also cast the PS in the following form
\beqra
&&
 P^{Zeld}(k)= e^{-k^2 \sigma_{v}^2 } \sum_{n=1}^\infty n! \int \frac{d^3p_1}{(2 \pi)^3}\cdots \frac{d^3p_n}{(2 \pi)^3}\,(2\pi)^3 \delta_D \left(\bk-\sum_{j=1}^n \bp_j\right)\nonumber\\
 &&\qquad\qquad\qquad\qquad \times  \left[F_n(\bp_1,\cdots,\bp_n) \right]^2 P^{lin}(p_1)\cdots P^{lin}(p_n)\,,
 \label{Zeld2}
\eeqra
where the kernel functions are given by
\beq
F_n(\bp_1,\cdots,\bp_n) =\frac{1}{n!}  \frac{\bk\cdot \bp_1}{p_1^2} \cdots \frac{\bk\cdot \bp_n}{p_n^2}\,.
\eeq
Isolating the first term in the sum \re{Zeld2}, it can be rewritten as
\beq
 P^{Zeld}(k)= e^{-k^2 \sigma_{v}^2 }  P^{lin}(k) + P_{MC}^{Zeld}(k)\,, 
 \label{Zeld3}
\eeq
showing explicitly that our eq.~\re{PS1} amounts to neglecting the mode-coupling contribution in the Zel'dovich approximation.

\begin{figure}
\centering{ 
\includegraphics[width=.45\textwidth,clip]{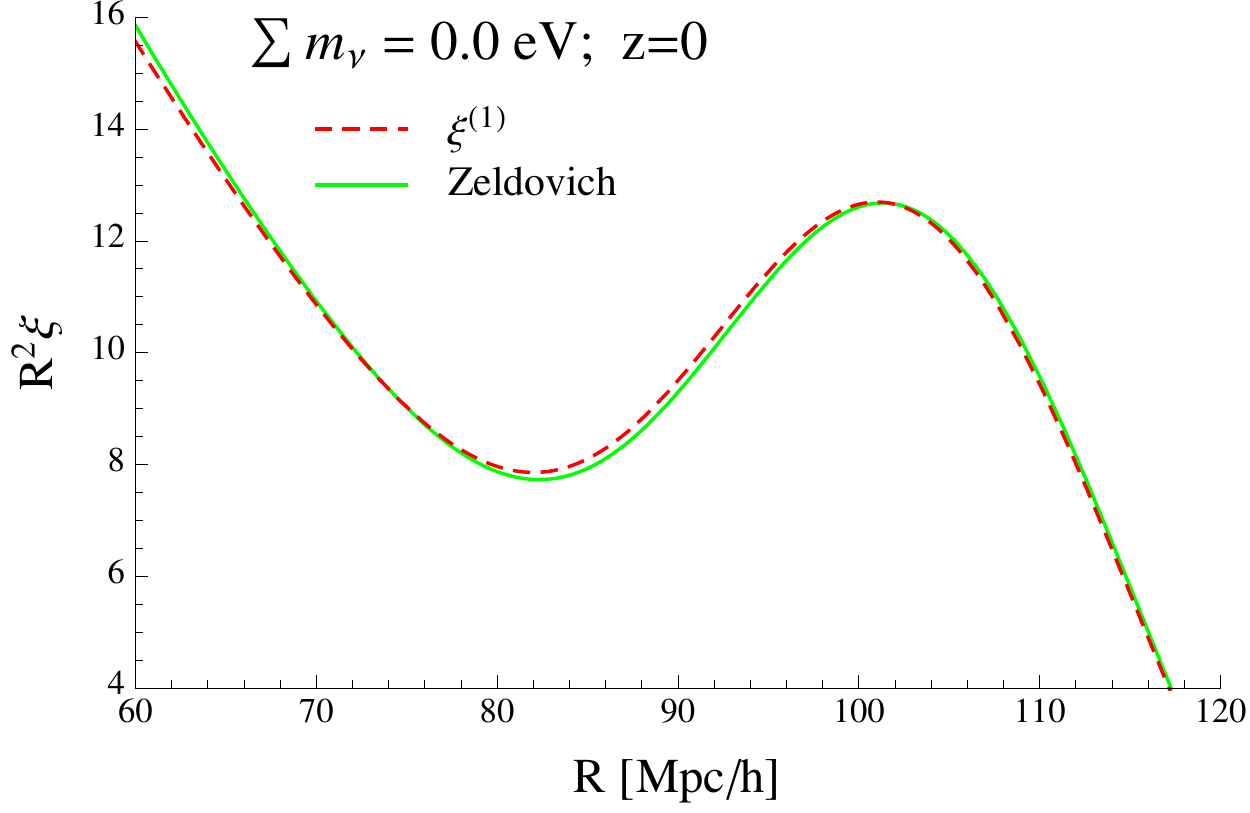}
}
\caption{CF for matter in real space, for massless neutrino cosmologies  and at redshift $z=0$. The green solid lines are obtained from  $\xi^{(1)}$ (based on the PS eq.~\re{PS1}, that is from the Zel'dovich approximation in which the mode-coupling contribution is neglected, see eq.~\re{Zeld3}), while the red dashed lines are obtained from the Zel'dovich approximation (based on the PS~\re{Zeldf}).
}
\label{fig:Zeldfig-abs}
\end{figure}

\begin{figure}
\centering{ 
\includegraphics[width=.45\textwidth,clip]{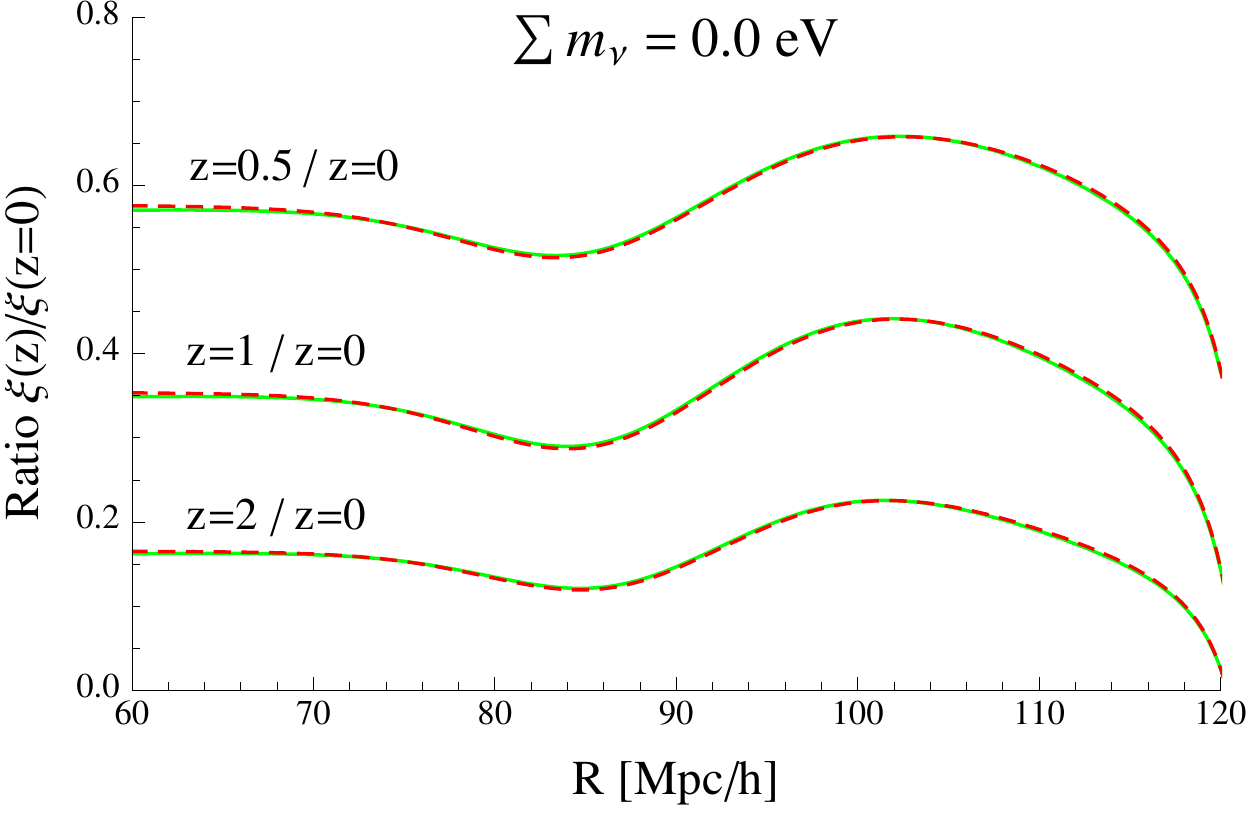}
\includegraphics[width=.45\textwidth,clip]{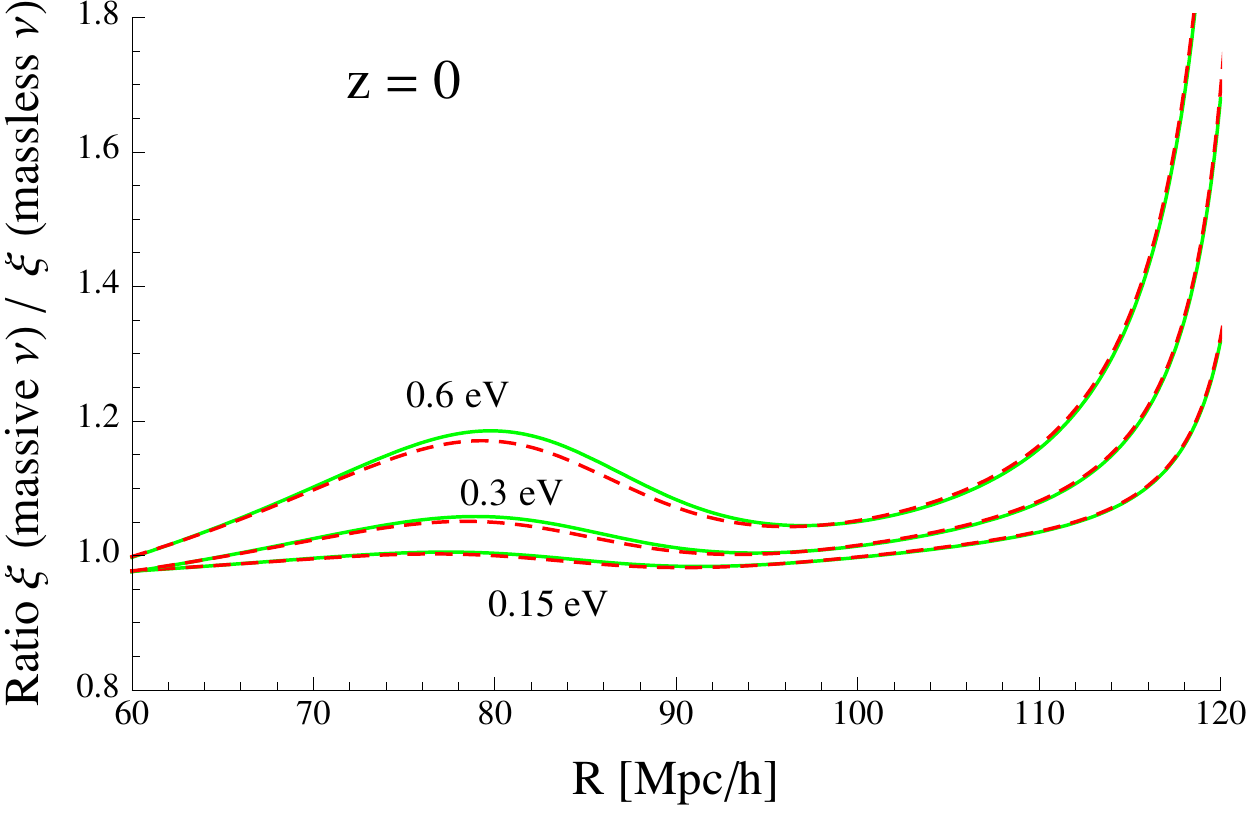}
}
\caption{Ratios of matter CFs in real space. On the left panel we show ratios of CFs at different redshifts, for massless neutrinos. On the right panel we show the $z=0$  CF for massive neutrinos (the label on the curve giving $\sum m_\nu$ in eV), divided by the corresponding CF for massless neutrinos.  The green solid lines are obtained from  $\xi^{(1)}$ (based on the PS eq.~\re{PS1}), while the red dashed lines are obtained from the Zeldovich approximation (based on the PS~\re{Zeldf}).
}
\label{fig:Zeldfig-rat}
\end{figure}

The comparison between the CF's obtained from eq.~\re{PS1} and from  the full Zel'dovich approximation, eq.~\re{Zeldf}, is shown in Figures~\ref{fig:Zeldfig-abs} and~\ref{fig:Zeldfig-rat}, from which one sees clearly that the mode-coupling part is completely negligible in the BAO peak region. Moreover, we also see that the two functions give essentially an identical result for ratios of correlation functions (both between different redshifts within one cosmology, and between different cosmologies with different neutrino masses).

\section{Redshift space}
\label{RSD}

We want to describe the effect of bulk flows on the CF or on the PS,  in redshift space. In real space, the position of a particle at time $\tau$ is given by
\beq
\bx(\tau) = \bx(0) + \int_0^\tau d\tau^\prime \bv(\tau^\prime)\,.
\eeq
In redshift space it is shifted to 
\beq
\bs(\tau) = \bx(\tau) + \frac{v^z(\tau)}{{\cal H}}\; \hat \bz\,,
\label{rsshift}
\eeq
where ${\cal{H}}$ is the Hubble parameter computed using conformal time, we work in the distant observer approximation and we set the line of sight along the z-axis. Now we split the particle velocity into a short and a long distance component,
\beq
\bv(\tau) = \bv_{sh}(\tau)+\bv_l(\tau)\,,
\eeq
so that 
\beq
\bs(\tau) = \by(\tau) + \frac{v_{sh}^z(\tau)}{{\cal H}}\; \hat \bz +  \frac{\bv_l(\tau)}{{\cal H}\,f} +  \frac{v_l^z(\tau)}{{\cal H}}\; \hat \bz\,,
\eeq
where
\beq
\by(\tau)=\bx(0) + \int_0^\tau d\tau^\prime \bv_{sh}(\tau^\prime)\,,
\eeq
and we have assumed that $\bv_l$ evolves according to linear PT.

Then, imposing mass conservation
\beq
(1+\delta_s(\bs))d^3 s = (1+\delta(\by))d^3 y\,,
\eeq 
and Fourier transforming, we have the density contrast in redshift space
\beq
\delta_D(\bk)+\delta_s(\bk) = \int d^3 x \, e^{-i \bk\cdot\bx} e^{-i k^z v^z_{sh} /{\cal H} }e^{-i \bv_l\cdot (\bk+f k_z {\bf \hat z})/({\cal H} f)}\,
(1+\delta(\bx))\,.
\eeq
The statistical average over long modes only gives
\beqra
&& \delta_D(\bk)+\langle \delta_s(\bk) \rangle_l = \int d^3 x \, e^{-i \bk\cdot\bx} e^{-i k^z v^z_{sh} /{\cal H} } \langle e^{-i \bv_l\cdot (\bk+f k_z {\bf \hat z})/({\cal H} f)}\rangle_l\,
(1+\delta(\bx))\nonumber \\
&& \qquad\qquad\qquad\;\;= \int d^3 x \, e^{-i \bk\cdot\bx} e^{-\frac{ k^2 \sigma_v^2 (1+\mu^2 f (2+ f)) }{2}}e^{-i k^z v^z_{sh} /{\cal H} } \,
(1+\delta(\bx))\,,
\label{delta}
\eeqra
where
$\mu=\hat \bk\cdot \hat {\bf z}$.

We now expand the third exponential in \re{delta} to linear order in $v^z_{sh}$ and then we get the correlator
\beq
\langle  \langle \delta_s(\bk) \rangle_l \langle \delta_s(-\bk) \rangle_l  \rangle_{sh} = P_{sh}^{\rm Kaiser}(k,\mu) e^{-k^2 \sigma_v^2 (1+\mu^2 f (2+ f)) }\,,
\eeq
where 
\beq
P_{sh}^{\rm Kaiser}(k,\mu) = (1 + \mu^2 f)^2 P^{lin}_{sh}(k),\,
\eeq
with $P^{lin}_{sh}(k)$ the linear PS for the short modes.

\section{Resolution tests}
\label{sec:resolution_tests}

Here we study the dependence of our results with respect to mass and force resolution. In order to check whether our results are converged against mass and force resolution, we have run two different sets of simulations for the massless neutrino cosmological model: 10 simulations with $384^3$ CDM particles and 10 simulations with $512^3$ CDM particles. All simulations have a box size equal to 1000 Mpc$/h$ and the softening length is set to $1/40$ of the mean linear inter-particle separation. 

\begin{figure}
\centering 
\includegraphics[width=.65\textwidth,clip]{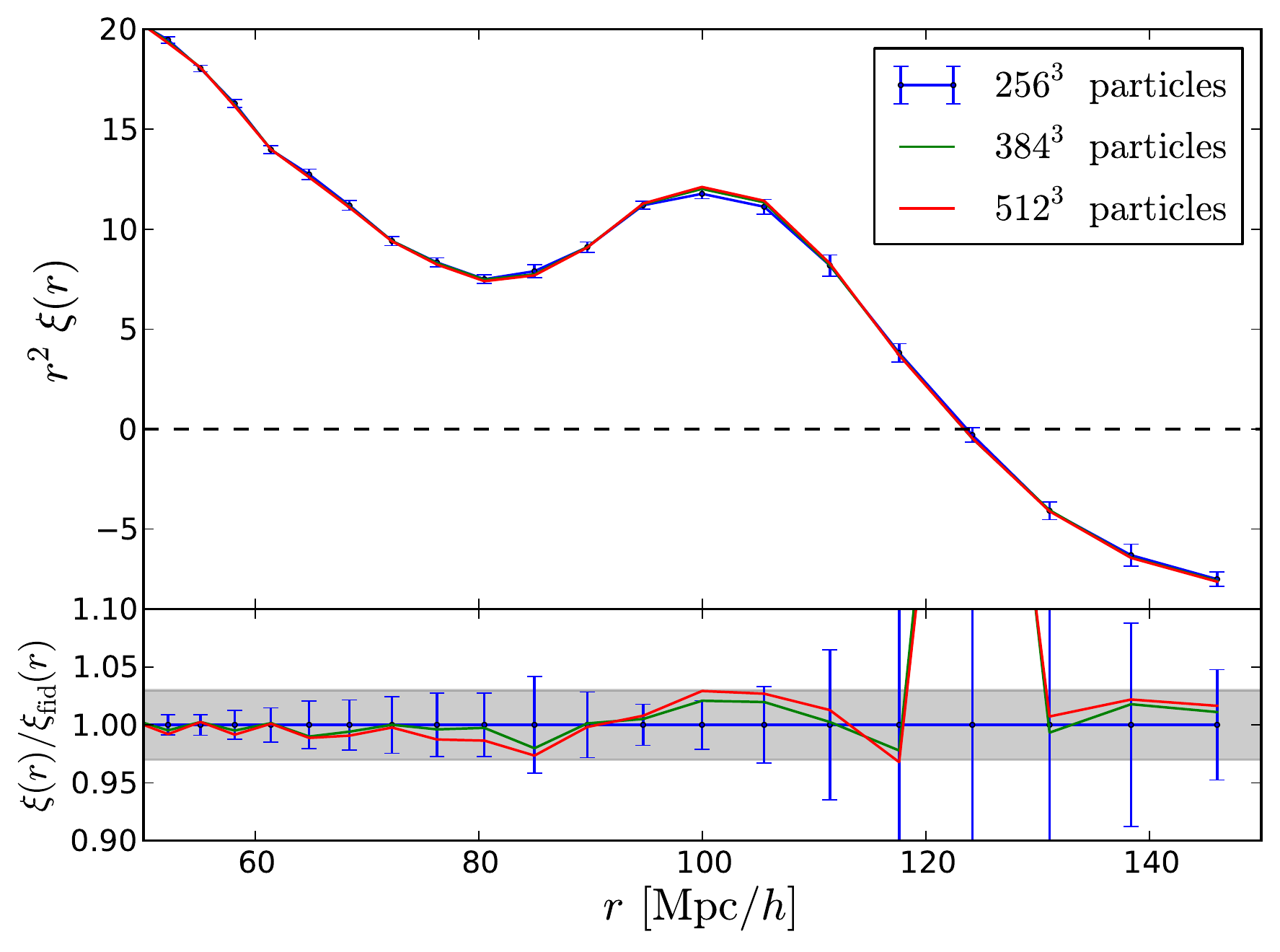}
\caption{\label{fig:CF_resolution} Two-point correlation function of the matter density field measured using 10 N-body simulations with $256^3$ (blue), $384^3$ (green) and $512^3$ (red) particles. The error bars represent the scatter around the mean value for the 10 realizations with $256^3$ particles divided by the square root of 10. The bottom panel displays the different correlation functions normalized to the fiducial one ($256^3$).}
\end{figure}

We have then measured the matter two-point correlation function in the different simulation sets and we show the results in Fig. \ref{fig:CF_resolution}. In that figure we also show the results obtained by using 10 realizations of our fiducial suite containing $256^3$ particles with the error bars representing the scatter around the mean value divided by $\sqrt{10}$.

We find that by increasing the simulation mass and force resolution the differences in the matter correlation function, in the range of scales investigated in this paper, are smaller than $\sim3\%$, and results from all simulations are compatible at $\sim1\sigma$.

 \section*{References}
\bibliographystyle{JHEP}
\bibliography{/Users/massimo/Bibliografia/mybib.bib}
\end{document}